\documentclass[namedreferences,hyperref,optionalrh,solaromanenum]{spr-sola}

\usepackage{graphicx}                    
\usepackage{color}                       

\usepackage[utf8]{inputenc}
\usepackage[T1]{fontenc}
\usepackage{booktabs}
\usepackage{float}
\usepackage{caption}
\makeatletter
\let\figure\@undefined \let\endfigure\@undefined
\newenvironment{figure}[1][tbp]{\@float{figure}[#1]}{\end@float}
\let\sola@tabular\relax
\makeatother

\usepackage{subcaption}
\usepackage{siunitx}
\usepackage{amsmath}
\usepackage{amssymb}
\usepackage{enumitem}

\let\cite\citep

\chardef\us=`\_

\begin{document}

\begin{frontmatter}

\title{Explainable Prediction of High-Speed Solar Wind Streams driven by Coronal Holes via Topological Data Analysis}

\author[addressref={aff1,aff2},corref,email={francesco.marchetti@unipd.it}]{\inits{F.}\fnm{Francesco}~\snm{Marchetti}}
\author[addressref={aff2,aff3},email={sabrina.guastavino@unige.it   }]{\inits{S.}\fnm{Sabrina}~\snm{Guastavino}}
\author[addressref={aff4,aff5},email={collin@gfz.de}]{\inits{D.}\fnm{Daniel}~\snm{Collin}}

\runningauthor{Marchetti, Guastavino and Collin}
\runningtitle{Explainable Prediction of HSSs via TDA}

\address[id=aff1]{Dipartimento di Matematica \lq\lq Tullio Levi-Civita\rq\rq, University of Padova, Italy}
\address[id=aff2]{Osservatorio Astrofisico di Torino, Istituto Nazionale di Astrofisica, Torino, Italy}
\address[id=aff3]{Dipartimento di Matematica, University of Genova, Italy}
\address[id=aff4]{Space Physics and Space Weather, GFZ Helmholtz Centre for Geosciences, Potsdam, Germany}
\address[id=aff5]{Department of Electrical Engineering and Computer Science, Technical University of Berlin, Berlin, Germany}
\begin{abstract}
High-speed solar wind streams (HSSs) originating from coronal holes (CHs) are an important driver of recurrent geomagnetic activity. Since HSSs are associated to coronal structures visible in extreme-ultraviolet observations, this provides a natural basis for forecasting several days in advance. In this work, we investigate whether the topology of coronal structures observed in extreme ultraviolet (EUV) images contains predictive information on the subsequent occurrence of CH-driven HSSs at Earth. To this end, we consider Topological Data Analysis (TDA), and in particular persistent homology, to classify EUV solar images according to the occurrence of HSSs at Earth four days later. Persistent homology is used to extract multiscale topological information from EUV images at wavelength 193~\AA, which is subsequently encoded through different vector representations and used as input to machine-learning classifiers. The target is a binary classification of HSS occurrence at a four-day lead time, defined using an HSS interval catalog and solar wind speed thresholds.
Several combinations of topological representations and classifiers are evaluated under a common cross-validation framework. The best-performing configurations, based on persistence-image representations and linear classifiers, achieve an area under the ROC curve (AUC) of approximately 0.82 and a True Skill Statistic (TSS) of approximately 0.43. These results indicate that the topology of EUV coronal structures alone contains predictive information on subsequent CH-driven HSS occurrence, despite the absence of explicit CH geometric descriptors in the model input. To investigate the origin of this predictive information, we develop a model-level explainability framework based on cross-validation-aggregated feature weights and spatial evidence maps. A fully traced example connects persistence diagrams and persistence-image features to the classifier output and to their spatial localization in the original EUV observation. The proposed framework provides a transparent and computationally lightweight approach for investigating the relationship between coronal topology and subsequent high-speed solar wind streams.
\end{abstract}

\begin{keyword}
\kwd{Solar wind}
\kwd{Coronal holes}
\kwd{Corona}
\kwd{Machine learning} \kwd{Topological Data Analysis} \kwd{High-speed solar wind streams}
\end{keyword}

\end{frontmatter}

\section{Introduction}
Near-Earth space weather disturbances are primarily associated with coronal mass ejections (CMEs; \citet{webb2012,georgoulis2024prediction}) and high-speed solar wind streams (HSSs), both of which can perturb the near-Earth environment and affect satellites and technological infrastructure \cite{schwenn2006,temmer2021,richardson2018}. While CMEs are responsible for many of the most intense geomagnetic storms, recurrent HSSs represent an important source of sustained geomagnetic activity, particularly during the declining and minimum phases of the solar cycle \cite{richardson2000,tsurutani2006,telloni2020study,guastavino2024forecasting}. HSSs are commonly associated with coronal holes (CHs), regions of reduced extreme-ultraviolet (EUV) and X-ray emission characterized by predominantly open magnetic-field configurations that allow plasma to escape into the heliosphere. The link between coronal holes and fast solar wind has a long observational history: the seminal identification of a CH as source of a high-velocity stream dates back to the Skylab era \cite{krieger1973}, and subsequent studies have investigated the relation between CH properties, high-speed streams, and recurrent geomagnetic activity \cite{nolte1977,sheeley1976,cranmer2002}.

The association between CHs and HSSs provides a natural basis for forecasting solar wind conditions at Earth. Since their solar sources can be observed before the associated streams reach 1~AU, EUV observations of the solar disk can provide predictive information several days in advance, with typical Sun-Earth propagation times of the order of a few days. Nevertheless, establishing a quantitative relationship between solar observations and subsequent in-situ solar wind conditions remains challenging. The solar wind speed measured at Earth does not depend on a single property of the source region \cite{hofmeister2018dependence}, and the propagation and evolution of streams through the heliosphere can modify their signatures before they reach 1 AU \cite{richardson2018,hofmeister2022area}. Moreover, approaches based on explicit CH identification introduce additional dependence on segmentation procedures and boundary definitions \cite{reiss2021observational,reiss2024community}, as well as on the
association between individual CHs and the corresponding
in-situ streams.

Several studies have therefore investigated the relationship between observable properties of CHs and the characteristics of the associated HSSs. Empirical approaches have related solar wind speed to quantities such as CH area and position on the solar disk \cite{vrsnak2007,rotter2012}, while more recent studies have exploited image-derived features and machine-learning techniques to capture more complex relationships between coronal observations and the solar wind \cite{bu2019,samara2022,collin2025}. Physics-based models provide a complementary route by modeling the magnetic configuration of the corona and the propagation of the solar wind through the heliosphere. More recently, physically informed and compact machine-learning representations have also shown that useful predictive skill can be obtained without necessarily resorting to increasingly complex architectures \cite{collin2025}.

However, a key open issue is that predictive performance does not automatically imply interpretability. In many image-based models, the link between learned internal representation and physically meaningful solar structures remains opaque.

Topological Data Analysis (TDA) offers a complementary approach for characterizing the structure of solar images through their topological properties. 
In particular, persistent homology characterizes how topological structures evolve as a dataset is examined across a sequence of filtration levels. When applied to images, it allows connected components and loops to be followed across intensity levels rather than describing structures at a single, predefined threshold. Their appearance and disappearance are summarized through persistence diagrams, providing a multiscale representation of the topology of the image. Persistent homology has well-established mathematical foundations, including algorithmic formulations and stability results \cite{edelsbrunner2002,zomorodian2004,cohensteiner2006,demarchi2022}, and its use in data analysis has grown substantially through the development of vectorized representations suitable for statistical and machine-learning methods \cite{otter2017,chazal2021}. From the perspective of solar EUV observations, this provides an appealing way of retaining information on the organization and persistence of image structures without reducing the input a priori to a small set of predefined CH descriptors.

In this work, we investigate whether topological information extracted from full-disk 193 \AA\ EUV solar images contains predictive information on the occurrence of HSSs at Earth four days later. Persistent homology is used to characterize the images, and different vectorizations of the resulting persistence diagrams are evaluated in combination with machine-learning classifiers under a common cross-validation framework. Importantly, the models are based exclusively on topological representations: no explicit CH segmentation, area, location, or other geometrical CH descriptors are provided as input. Our objective is therefore not to introduce persistent homology as a new mathematical methodology, nor to propose a state-of-the-art operational HSS forecasting system, but rather to assess how much predictive information can be extracted from the topology of EUV observations alone.

A second objective is to investigate where the predictive information identified by the classifiers is located in the original solar observations. To this end, we complement the classification analysis with an explainability framework based on fold-aggregated model weights and spatial evidence maps. For a representative example, we trace the complete chain from the EUV image and its persistence diagram to the vectorized topological representation, standardized classifier input, linear model output, and spatially localized positive and negative evidence. This provides a direct connection between the mathematical topological representation, the resulting classification, and the solar structures contributing to the model decision.

The manuscript is structured as follows. Section \ref{sec:homology} introduces persistent homology with emphasis on the concepts needed in this work. Section \ref{sec:pipeline} details the end-to-end pipeline and implementation choices. Section \ref{section:results_discussion} presents results and interpretability analysis, including a fully traced worked example. Conclusions are offered in Section \ref{sec:conclusions}.

\section{On persistent homology}\label{sec:homology}

Topological Data Analysis (TDA) provides a mathematical framework for characterizing the structure of data through their topological properties. In this work, we use \emph{persistent homology}, a TDA technique that describes how topological structures appear, evolve, and disappear as a scalar field is examined across multiple levels \cite{edelsbrunner2002,zomorodian2004,otter2017}. When applied to an image, persistent homology therefore provides a multiscale description of its intensity structure without requiring the selection of a single threshold.

\subsection{From an image to a persistence diagram}

Let $I(u,v)\in\mathbb{R}$, $(u,v)\in\mathbb{N}\times\mathbb{N}$
denote the intensity associated with pixel $(u,v)$ in a two-dimensional image $I$. The pixel grid can be analyzed through a cubical complex, in which pixels and their adjacencies define the underlying combinatorial structure \cite{otter2017,chazal2021}.

\begin{figure}[H]
\centering
\begin{subfigure}{0.31\textwidth}
\centering
\includegraphics[width=\textwidth]{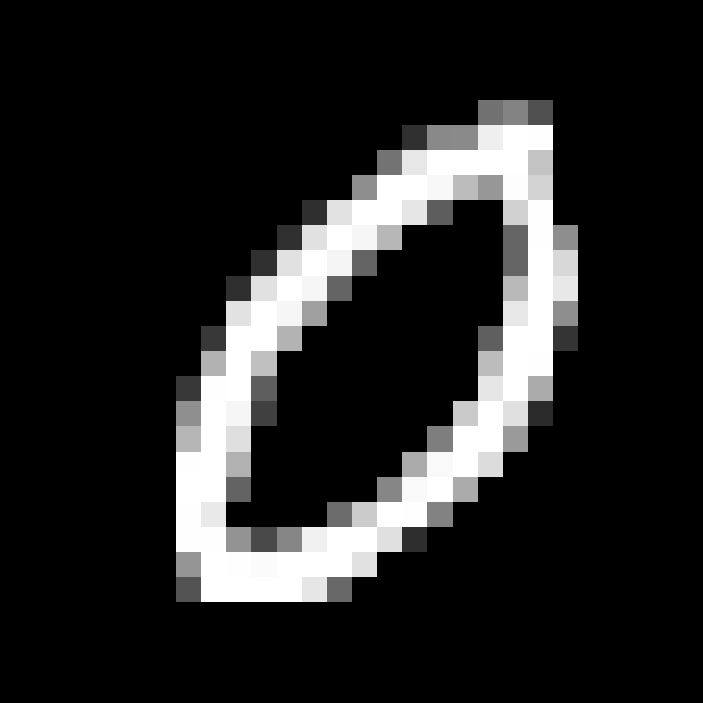}
\caption{A ``0'' from MNIST.}
\end{subfigure}
\begin{subfigure}{0.31\textwidth}
\centering
\includegraphics[width=\textwidth]{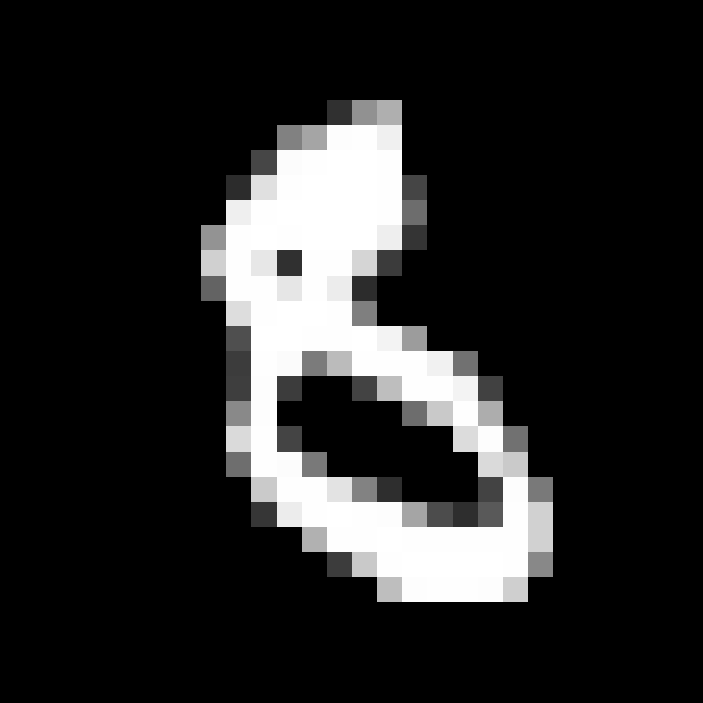}
\caption{An ``8'' from MNIST.}
\end{subfigure}
\caption{MNIST images considered.}
\label{fig:zero_e_otto}
\end{figure}

A simple way of investigating the structure of an image is to select an intensity threshold $\alpha$ and consider only pixels whose intensities lie above or below that threshold. For example, two standard choices consist in taking sublevel or superlevel set defined as
\begin{equation}\label{eq:filtrations}
    K^{\mathrm{sub}}_\alpha = \mathcal{K}(\{(u,v)\mid I(u,v)\le \alpha\}),
\qquad
K^{\mathrm{sup}}_\alpha = \mathcal{K}(\{(u,v)\mid I(u,v)\ge \alpha\}),
\end{equation}
where $\mathcal{K}(\cdot)$ denotes the induced complex. In practice, both conventions are equivalent up to monotone transformation (e.g., replacing $I$ by $\max(I)-I$).
In the following introduction, we refer to the superlevel approach because it provides a direct visualization of the thresholding process with the considered examples, which are a ``0'' and ``8'' digit from the well known MNIST handwritten-digit dataset \cite{lecun1998} (see Figure \ref{fig:zero_e_otto}). 

At each filtration level, homology characterizes topological structures of different dimensions. For the two-dimensional images considered here, we focus on $H_0$,
which describes connected components, and
$H_1$
which describes one-dimensional holes or loops. Formally, the $h$-th homology group identifies $h$-dimensional cycles that are not boundaries; a detailed algebraic treatment can be found, for example, in \cite{edelsbrunner2002,zomorodian2004,chazal2021}. For the purposes of the present work, the relevant point is that $H_0$ and $H_1$ provide complementary descriptions of the connectivity and hole structure of image-intensity patterns.

\begin{figure}[H]
	\centering
	\begin{subfigure}{0.31\textwidth}
		\centering
		\includegraphics[width=\textwidth]{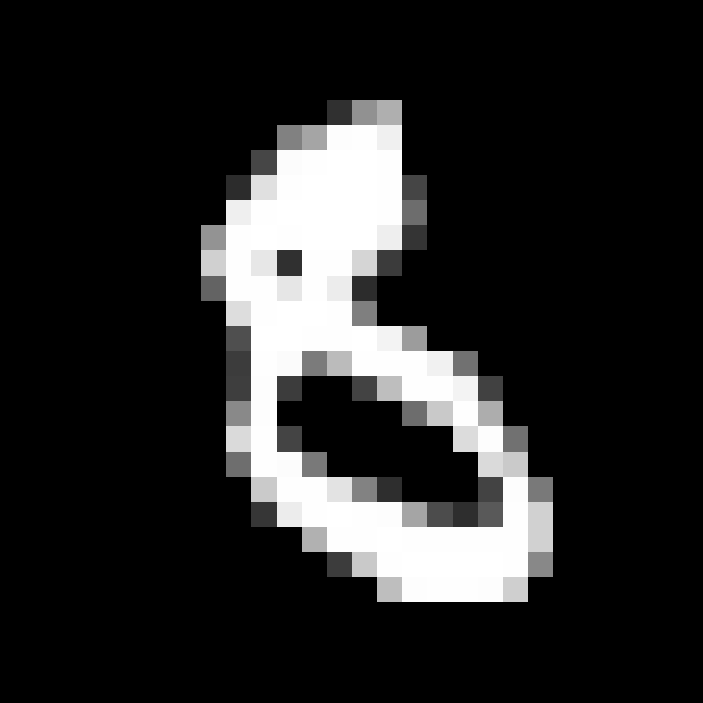}
		\caption{MNIST 8 original image.}
	\end{subfigure}
	\hfill
	\begin{subfigure}{0.31\textwidth}
		\centering
		\includegraphics[width=\textwidth]{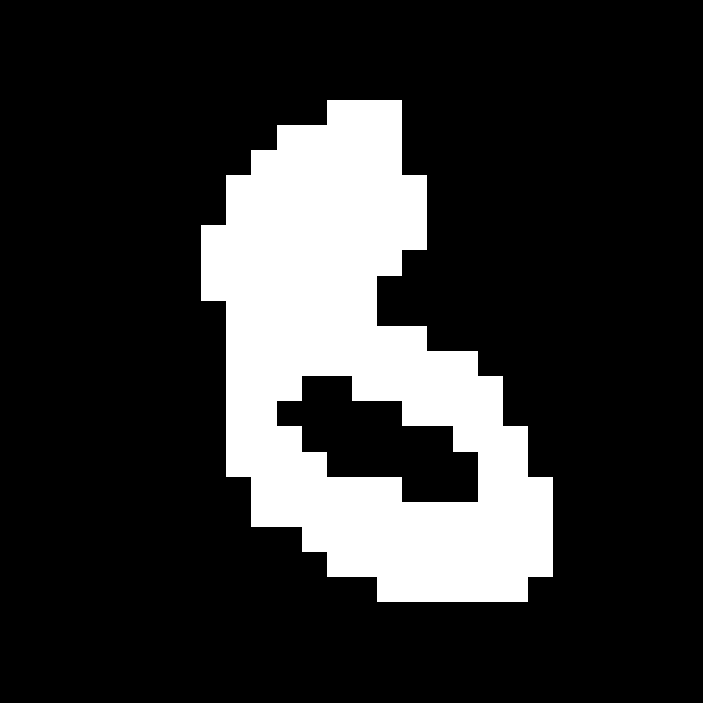}
		\caption{Superlevel set at $\alpha=0.166$.}
	\end{subfigure}
	\hfill
	\begin{subfigure}{0.31\textwidth}
		\centering
		\includegraphics[width=\textwidth]{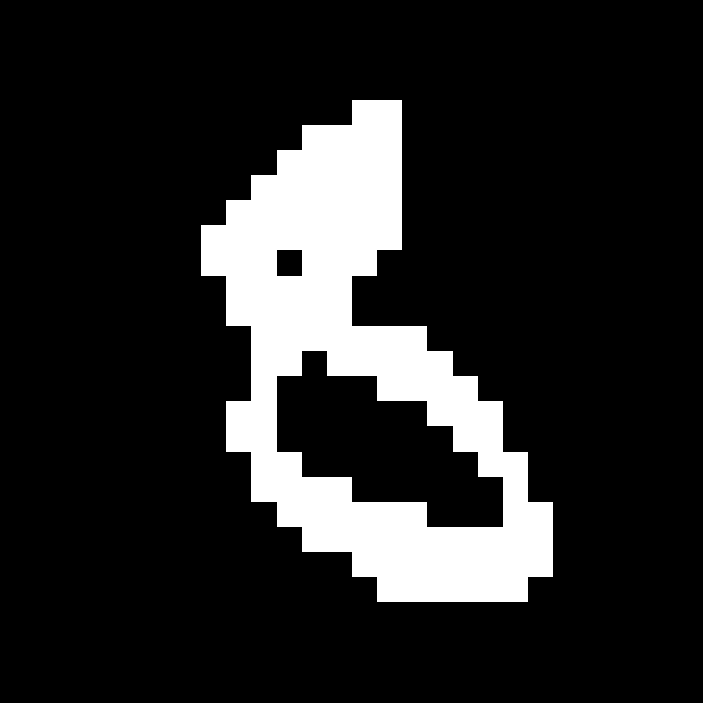}
		\caption{Superlevel set at $\alpha=0.519$.}
	\end{subfigure}
	\caption{MNIST-8 example.}
	\label{fig:mnist_sublevels}
\end{figure}

As $\alpha$ varies, the topology of $K^{\mathrm{sub}}_\alpha$ or $K^{\mathrm{sup}}_\alpha$ changes: connected regions can appear or merge, and holes can form or disappear. The dependence of these structures on the selected threshold can be illustrated using a simple example. Figure \ref{fig:mnist_sublevels} shows two superlevel sets obtained at two different intensity thresholds from the ``8'' digit. The topology of the active set $\{(u,v)\:|\:I(u,v)>\alpha\}$ changes as the threshold varies: structures that are visible at one level may merge or disappear at another level. In particular, one snapshot has one hole, while another has two holes. Consequently, describing the image at a single threshold would retain only a particular snapshot of its topology. 

Rather than selecting a single threshold, persistent homology considers the entire sequence of thresholded sets, referred to as a \emph{filtration}.
Sublevel sets grow with $\alpha$ ($K^{\mathrm{sub}}_\alpha\subseteq K^{\mathrm{sub}}_\beta$ for $\alpha\le\beta$), while superlevel sets shrink. Therefore, each topological structure is tracked throughout the filtration: A structure is assigned a \emph{birth} value $b$, corresponding to the filtration level at which it appears, and a \emph{death} value $d$, corresponding to the level at which it disappears or merges into an older structure, building a persistence pair
$(b,d)$, $b<d$, with persistence $p=d-b$.
The collection of persistence pairs in homology dimension $h$ defines the \emph{persistence diagram}
$D[h]=\{(b_i,d_i)\}_{i=1}^{N_h}$, where $N_h$ denotes the number of topological generators in homology dimension $h$.
Points farther from the diagonal $b=d$ correspond to structures that persist over a larger range of filtration values, whereas points close to the diagonal represent short-lived topological structures and are generally more sensitive to perturbations. Persistence itself, however, does not determine whether a structure is relevant for a particular prediction problem: short- and long-lived generators may carry different amounts of information depending on the data and task.

\begin{figure}[H]
\centering
\begin{subfigure}{0.31\textwidth}
\centering
\includegraphics[width=\textwidth]{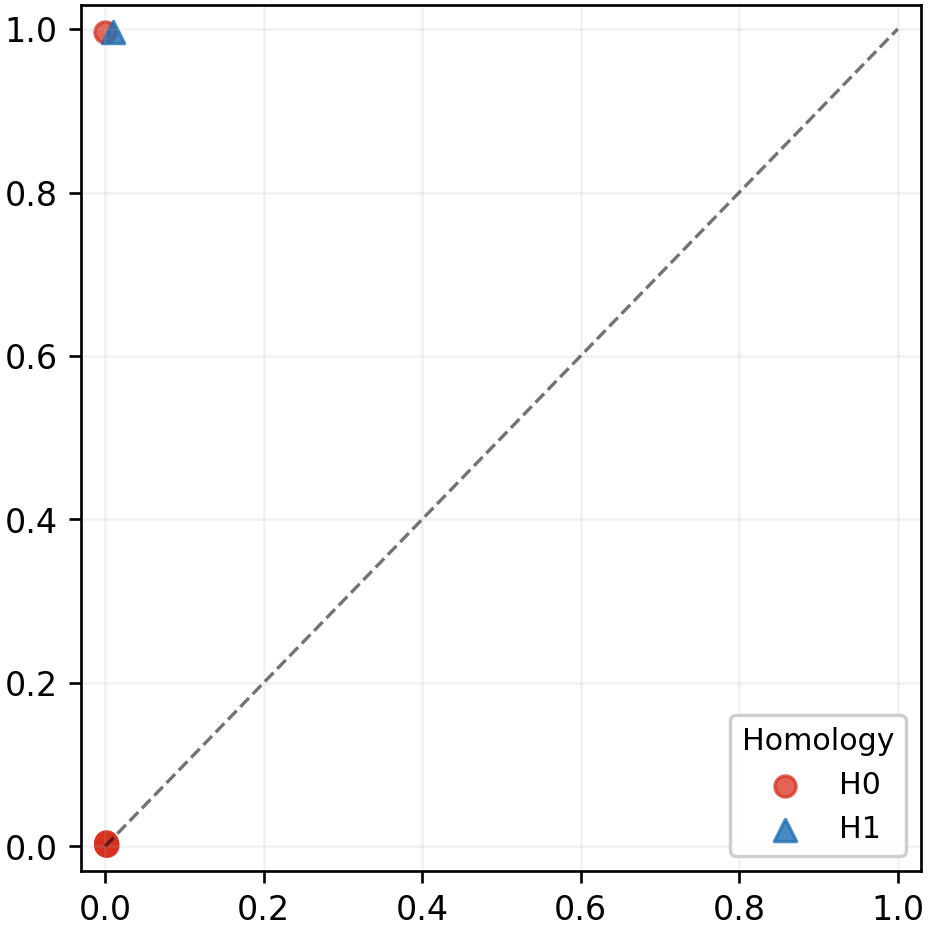}
\caption{Persistence diagram extracted from Figure \ref{fig:zero_e_otto}a.}
\end{subfigure}
\begin{subfigure}{0.31\textwidth}
\centering
\includegraphics[width=\textwidth]{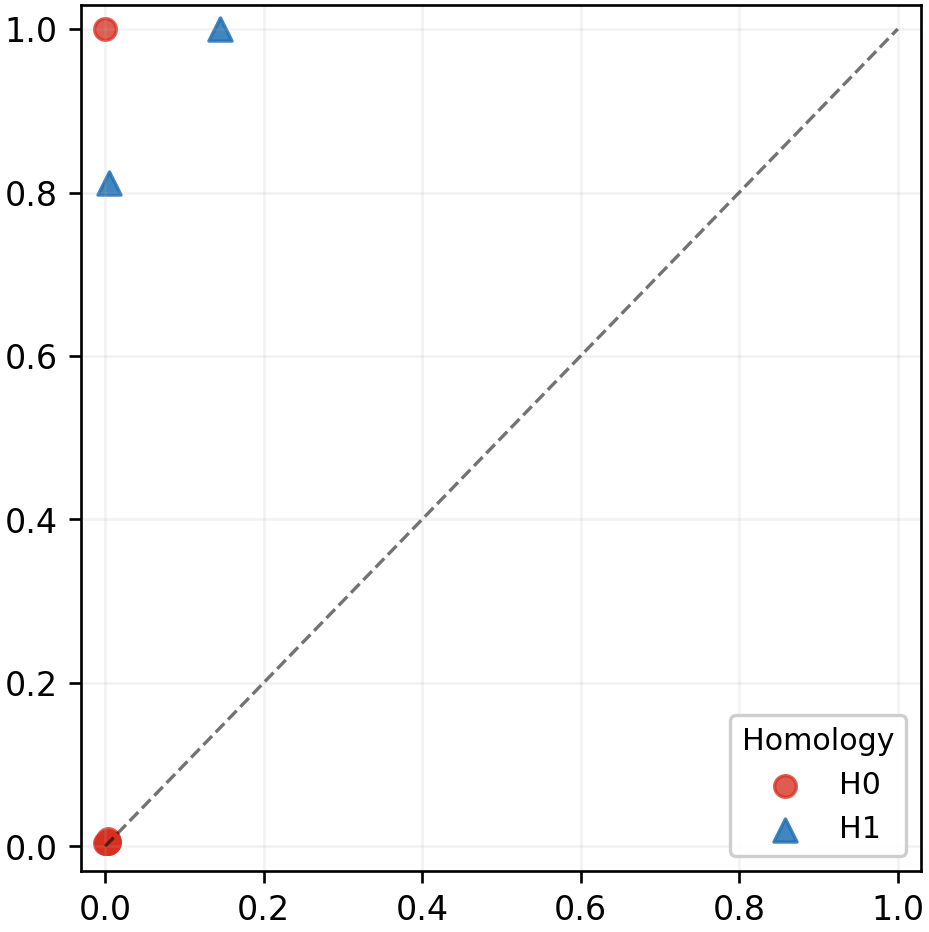}
\caption{Persistence diagram extracted from Figure \ref{fig:zero_e_otto}b.}
\end{subfigure}
\caption{Persistence diagrams constructed from Figure \ref{fig:zero_e_otto}. Both $H_0$/$H_1$ homologies are shown. Birth and death coordinates refer to the $x$- and $y$-axis, respectively. We find important generators far from the diagonal, whereas points close to the diagonal yield unstable, noisy generators.}
\label{fig:persistence_diagrams}
\end{figure}

To visualize this construction, we compute the persistence diagrams of the two MNIST samples in Figure \ref{fig:persistence_diagrams}. We observe that the digit ``0'' yields one dominant $H_0$ generator (one connected component) and one dominant $H_1$ generator (one principal hole), while the digit ``8'' yields one dominant $H_0$ generator and two dominant $H_1$ generators (two principal holes). 

An important mathematical property of persistence diagrams is their stability with respect to perturbations of the underlying scalar field \cite{cohensteiner2006}, that is, controlled perturbations of image intensities induce controlled changes in their persistence diagrams. This property provides an important motivation for using persistent homology as a representation of image structure, being suitable in case of noisy data, although its practical behavior also depends on the adopted preprocessing and filtration.


\subsection{Vectorizing persistence diagrams for machine learning}

Even if persistence diagrams provide a compact representation of topological structures, they are multisets of birth--death pairs whose cardinality generally varies from image to image. Therefore, a persistence diagram
cannot be directly used as a fixed-dimensional input to standard machine-learning algorithms. To address this issue, a \emph{vectorization} maps $D[h]$ into a fixed-dimensional representation $x[h]\in\mathbb{R}^d$ that can subsequently be used as input to a machine-learning model \cite{otter2017,chazal2021}.

In this work, we consider four commonly used persistence-based representations: Betti curves, persistence landscapes, silhouettes, and persistence images. These representations retain different aspects of the information contained in the persistence diagram and therefore allow us to investigate whether the predictive information depends on how the underlying topology is encoded.

\textbf{Betti curves.} The Betti curve counts how many classes are alive at each filtration level $t$:
\[
\beta_k(t)=\sum_i \mathbf{1}_{\{b_i\le t<d_i\}}.
\]
Sampling $\beta_k(t)$ on a fixed grid yields a finite-dimensional vector. Intuitively, this captures \emph{how much topology is active} at each scale, without tracking individual generators \cite{otter2017,chazal2021}. In the context of EUV solar images, the Betti curve describes how the number of topological structures changes as different intensity levels are explored. For $H_0$, it summarizes the number of disconnected intensity regions present at each filtration level, whereas for $H_1$ it describes the number of holes enclosed by such regions. Therefore, it provides a global description of how the connectivity of the EUV intensity pattern evolves across the filtration, without retaining the birth and death coordinates of individual generators.

\textbf{Persistence landscapes.} For each point $(b_i,d_i)$, define the triangular (tent) function
\[
\lambda_i(t)=\max\!\big(0,\min(t-b_i,d_i-t)\big).
\]
This is zero outside $[b_i,d_i]$, peaks at $(b_i+d_i)/2$, and has height $(d_i-b_i)/2$. A \emph{landscape layer} $\Lambda_m(t)$ is the $m$-th largest value among $\{\lambda_i(t)\}_i$ at each $t$. For an EUV image, persistence landscapes describe at which intensity-related filtration levels the most persistent topological structures are present and how long they remain across the filtration. The first landscape layers (low $m$) emphasize the dominant persistent structures, while higher layers (large $m$) retain information about additional structures occurring over similar filtration ranges. They therefore provide a representation of the hierarchy of persistent connectivity and hole patterns present in the solar image \cite{bubenik2015}.

\textbf{Silhouettes.} A silhouette is a weighted average of landscape functions:
\[
\mathrm{Sil}[h](t)=\frac{\sum_i \omega_i\lambda_i(t)}{\sum_i \omega_i},
\qquad
\omega_i=(d_i-b_i)^p\;(p\ge 0).
\]
Compared with full landscapes, silhouettes compress information into one curve while emphasizing persistent generators through $w_i$ \cite{chazal2021}. In an EUV solar image, the silhouette provides a global persistence-weighted profile of the topological structures encountered across the intensity filtration. Rather than retaining separate information for individual generators or landscape layers, it summarizes where in the filtration the overall topological activity is concentrated. The parameter $p$ controls the emphasis placed on long-lived structures: increasing $p$ gives progressively greater influence to structures that persist across a wider range of intensity-related filtration levels.

\textbf{Persistence images.} First map points to birth-persistence coordinates\\ $(b_i,p_i)$ with $p_i=d_i-b_i$. Then place weighted Gaussian kernels and integrate on a fixed grid:
\begin{equation}\label{eq:pers_image}
\rho[h](x,y)=\sum_i w(p_i)\,\exp\!\left(-\frac{(x-b_i)^2+(y-p_i)^2}{2\sigma^2}\right),
\end{equation}
followed by pixel-wise integration. Flattening the resulting grid gives a vector of fixed length. This representation is thus stable and naturally compatible with linear models \cite{adams2017}. For an EUV solar image, different regions of the persistence image therefore correspond to topological structures emerging at different intensity-related levels and characterized by different lifetimes. In particular, the representation distinguishes short-lived structures from structures that remain topologically identifiable over a broad range of the filtration. By retaining this two-dimensional organization, persistence images make it possible to identify which ranges of birth and persistence are most relevant to the classifier and, subsequently, to relate these contributions back to structures in the original solar image

The four representations therefore summarize persistence information in different ways: Betti curves encode the number of simultaneously active generators, landscapes retain ordered persistence profiles, silhouettes provide a weighted functional summary, and persistence images preserve a two-dimensional representation of the distribution of generators in birth--persistence space. Their practical behavior depends on choices such as filtration, preprocessing, weighting, smoothing, and discretization resolution \cite{cohensteiner2006,fasy2014}.

In the next Section \ref{sec:pipeline}, we describe how this persistent homology pipeline is constructed from the 193 \AA\ EUV observations and then used as input to the classification models. For the persistence-image representation, the connection between the original solar image, persistence diagram, vectorized features, classifier coefficients, and spatial evidence is further examined in Section \ref{sec:working_example}.

\section{Data and methodology}\label{sec:pipeline}

The proposed framework investigates whether the topology of full-disk EUV solar images contains information about the occurrence of high-speed solar wind streams (HSSs) at Earth four days later. The analysis consists of four main steps: i) definition of the image-HSS pairs and of the corresponding binary target; ii) preprocessing of the EUV observations and extraction of persistent-homology features; iii) classification using different vector representations and machine-learning models; and iv) interpretation of the learned topological features and their localization in the original solar images.

\subsection{Data and label definition}\label{sec:data_label}

Following the observational setting adopted in \cite{collin2025}, we combine three synchronized data sources. 
Full-disk extreme-ultraviolet (EUV) observations are obtained from the 193~\AA\ channel of the Atmospheric Imaging Assembly (AIA; \cite{lemen2012}) on board the Solar Dynamics Observatory (SDO; \cite{pesnell2012solar}), with an hourly cadence.
The images are corrected for position, rotation, and scaling of the solar disk, light scattering, and instrument degradation as described in \cite{collin2026}.
Near-Earth solar wind speed measurements are obtained from the hourly OMNI dataset \cite{papitashvili2020}, while HSS intervals are identified using the catalog constructed through the peak-filtering procedure described in \cite{collin2025,collin2026}.

The AIA 193~\AA\ channel is particularly suitable for the present analysis because coronal holes (CHs), which constitute the main source regions of HSSs, appear as regions of reduced EUV emission relative to the surrounding corona. Importantly, however, no explicit CH segmentation is performed in the proposed approach. The entire full-disk EUV image is used as the scalar field from which the topological representation is extracted.

For each EUV image acquired at time $t$, the classification target is defined according to the solar wind conditions observed at $t+4$ days. The four-day lead time is motivated by the typical transit time of CH-driven high-speed solar wind from the Sun to Earth and is consistent with the forecasting setup adopted in \cite{collin2025,collin2026}. The binary labels are defined as
\begin{equation*}
y =
\begin{cases}
1, & \begin{aligned}
     &\text{if the target time belongs to an HSS interval} \\
     &\text{with } \tilde{V}(t+4\,\mathrm{d}) > 500~\mathrm{km\,s}^{-1},
     \end{aligned} \\[3ex]
0, & \begin{aligned}
     &\text{if the target time does not belong to an HSS interval} \\
     &\text{and } \tilde{V}(t+4\,\mathrm{d}) < 400~\mathrm{km\,s}^{-1}.
     \end{aligned}
\end{cases}
\end{equation*}
Samples that do not satisfy either condition are excluded from the analysis. Here, $\tilde V(t+4\,\mathrm{d})$ denotes the OMNI hourly solar wind speed at the timestamp nearest to $t+4$ days (accepted only if the nearest sample is within 12 hours). Therefore, the threshold is applied to a single target-time speed value, not to a maximum over a time interval.


This definition deliberately introduces a separation between the positive and negative classes. The intermediate cases are excluded to reduce ambiguity in the association between the solar observation and the subsequent solar wind regime.

The resulting dataset contains $71{,}976$ labeled EUV images acquired between 2010-05-18 and 2024-06-30, of which $13{,}968$ are assigned to the positive ($y=1$) HSS class and $58{,}008$ to the negative ($y=0$) non-HSS class.

We point out that an important issue in the evaluation of HSS prediction models is the strong temporal dependence of the solar wind time series, which exhibits an auto-correlation for up to 4 days and recurring auto-correlation peaks every multiple of 27 days. The latter are associated with long-lived coronal holes that can produce recurrent HSSs reaching Earth over successive solar rotations.
A random CV split would cause data leakage if either temporally close data points or recurring coronal holes appear in both training and test data. To avoid this, we discard 90 days of data between training, validation, and test datasets.

\begin{figure}[H]
\centering
\includegraphics[width=0.7\textwidth]{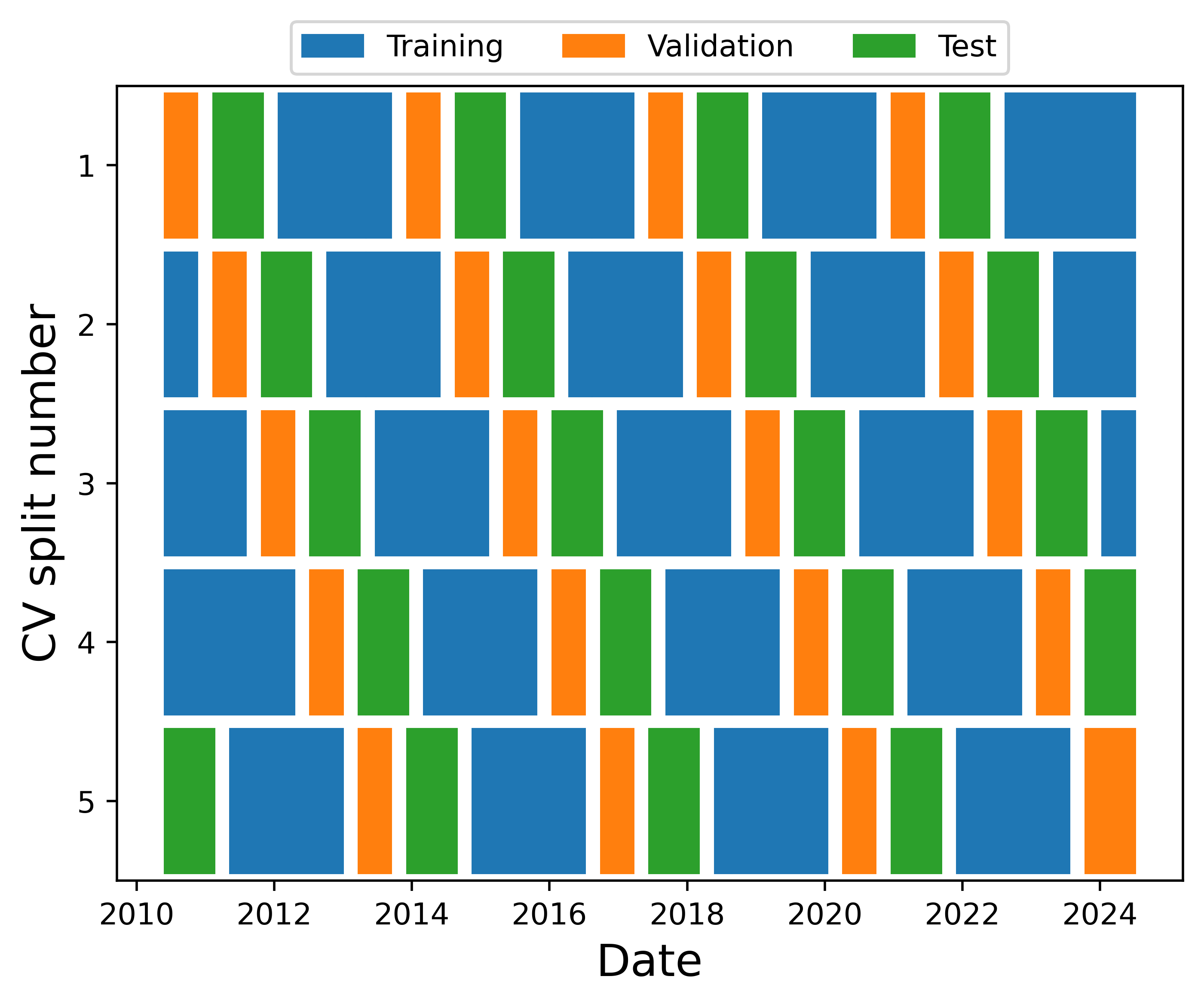}
\caption{Chronological 5-fold split used in this work. Colors indicate train/test allocation by fold over time; each test fold is a contiguous temporal block to reduce recurrence-driven leakage.}
\label{fig:cv_split}
\end{figure}

Furthermore, the state of the solar corona and correspondingly the solar wind parameters vary over the solar cycle \cite{richardson2000,tsurutani2006}. 
Therefore, we ensure that data subsets are distributed across the solar cycle. To form the five subsets, we divide the dataset into 20 continuous chunks of approximately nine months each, and assign to each subset four of those chunks, while ensuring that each data point appears exactly once as test data. 
The resulting CV splits can be seen in Figure~\ref{fig:cv_split}. The overall performance of a tested model is assessed by computing the evaluation metrics on the aggregated set of all test data predictions.

\subsection{Preprocessing and topology extraction assumptions}

Since the images have values spanning different orders of magnitude, we normalize them as suggested by \cite{jarolim2021}. First, we clip the values to an upper threshold of 7757.31. Then, we rescale the values to $[0,1]$, apply an inverse hyperbolic sine function stretch to the image $I$ as
\[
I_{\mathrm{norm}}= \frac{\mathrm{asinh}(I/a)}{\mathrm{asinh}(1/a)}, \qquad a=0.005,
\]
and rescale to $[-1,1]$. Here, $a>0$ controls contrast in low-intensity regions. These steps enhance coronal-hole–specific features, reduce strong variations near active regions, and ensure a consistent numerical range.

Further, we invert the normalized images, using
\[
I_{\mathrm{proc}} = \max(I_{\mathrm{norm}}) - I_{\mathrm{norm}}.
\]
In the pipeline, unlike the examples in the previous section, we then use sublevel filtration on the processed image $I_{\mathrm{proc}}$. The inversion has a practical role: CH regions are dark in raw EUV intensity, and inversion turns them into peaks in $I_{\mathrm{proc}}$, so CH-like structures are entered early and consistently in the sweep. Moreover, inversion changes which homology classes carry the physically relevant signal. Without inversion, CH darkness is represented indirectly through complements/background, and $H_0$ fragment statistics become sensitive to weak contrast fluctuations. With inversion, CH boundaries are more naturally represented by loop-like structures, and robust $H_1$ generators associated with CH morphology are separated earlier from near-diagonal noise. Importantly, we do not assume an a priori correspondence between coronal holes and a specific homology dimension. The $H_0$ and $H_1$ persistence diagrams, respectively, characterize the evolution of connected components and hole-like structures across the filtration. Their relevance to HSS classification, and their spatial association with structures in the original EUV images, are subsequently investigated through the classification and explainability analyses.

\subsection{Classification models and evaluation}

As previously described, a persistence diagram $D[h]$ built on a preprocessed EUV image $I_{\mathrm{proc}}$ is vectorized in $\mathbb{R}^d$ as $x[h]$, $h\in\{0,1\}$. The contribution of both homologies can be stacked in a vector
\begin{equation*}
    x = (x[0],x[1])\in\mathbb{R}^{2d}.
\end{equation*}
Let $x=(x_1,\ldots,x_{2d})$ denote the components of this resulting topological features vector. First, we standardize each feature as
\begin{equation*}
x_j^{\mathrm{std}}=\frac{x_j-\mu_j}{\sigma_j},
\qquad j=1,\ldots,2d,
\end{equation*}
where $\mu_j$ and $\sigma_j$ are the mean and standard deviation of feature $j$ computed on the training subset of the corresponding fold. This distinction is important for interpretation: $x$ retains the original topological amplitudes, whereas $x^{\mathrm{std}}$ is the representation on which the classifier will be fitted and evaluated.

Since different classifiers may exploit the information encoded in the topological features in different ways, we consider a set of models spanning linear, tree-based, boosting, and kernel-based approaches. This allows us to investigate not only whether the topological representations contain predictive information, but also whether such information can be captured through relatively simple relationships or requires more complex nonlinear decision rules.

\textbf{Logistic Regression and LinearSVM.}
We first consider two linear classifiers, Logistic Regression and a linear Support Vector Machine (LinearSVM) \cite{cox1958,cortes1995}. Logistic Regression models the probability of HSS occurrence as
\begin{equation}\label{eq:linear_model}
p(y=1\mid x)=\textrm{sigmoid}(w^\top x+\delta),
\qquad
\textrm{sigmoid}(z)=\frac{1}{1+e^{-z}},
\end{equation}
where $w$ contains the coefficients associated with the topological features and $\delta$ is the intercept. LinearSVM similarly separates the two classes through a linear decision boundary, but determines this boundary by maximizing the separation margin between the classes. These models provide a useful baseline for assessing whether HSS and non-HSS observations can already be distinguished through linear combinations of the topological features. Moreover, their signed coefficients provide a direct connection between individual features and the model decision, making linear models particularly suitable for the explainability analysis presented in Section \ref{sec:xai}.

\textbf{Random Forest and Extra Trees.}
Random Forest and Extra Trees are ensemble methods that combine the predictions of multiple decision trees \cite{breiman2001,geurts2006}. Unlike the linear models, they can capture nonlinear relationships and interactions among the topological features without requiring these relationships to be specified in advance. Random Forest introduces variability among the individual trees through resampling of the training data and random feature selection, whereas Extra Trees further increases randomization when constructing the decision boundaries within individual trees. Their inclusion therefore allows us to assess whether nonlinear combinations of topological features provide additional information for distinguishing images associated with subsequent HSS occurrence.

\textbf{XGBoost.}
XGBoost is a gradient-boosting approach in which decision trees are constructed sequentially, with each new tree aimed at improving the errors of the current ensemble \cite{chen2016}. This iterative construction makes it possible to represent complex nonlinear relationships and feature interactions while controlling model complexity through regularization. In the present analysis, XGBoost provides a complementary test of whether combinations of topological features that are not captured by linear decision boundaries improve the classification of HSS and non-HSS observations.

\textbf{GreedyMatern.}
Finally, we consider a kernel-based classifier employing a Matérn kernel together with a greedy selection strategy. Kernel methods implicitly represent the observations in a feature space in which nonlinear relationships between the original topological descriptors can be captured through pairwise similarities. The Matérn kernel provides a flexible notion of similarity controlled by its smoothness and length-scale parameters, while the greedy procedure selects a reduced set of representative basis elements. This provides a nonlinear alternative to the tree-based approaches while retaining a comparatively compact model representation \cite{camattari2024,rasmussen2006,wenzel2023}.

Taken together, these models cover substantially different assumptions on the relationship between the topological representation and the HSS target. The linear classifiers test whether the relevant information can be recovered through additive linear contributions of the topological features, whereas the tree-based, boosting, and kernel approaches allow progressively more flexible nonlinear relationships. The comparison is therefore intended primarily to assess how consistently predictive information can be extracted from the different topological representations, rather than to identify an optimized operational forecasting architecture.

For each combination of topological vectorization and classifier, model selection and performance evaluation are carried out independently within each of the five temporal cross-validation folds described in Section \ref{sec:data_label}. Model hyperparameters are selected using only the training and validation subsets of the corresponding fold via a grid search. Once the model configuration has been selected, its performance is evaluated on the test subset, which is not involved in either model fitting or hyperparameter selection. Since the five test subsets are mutually exclusive, the predictions obtained across the five folds can be concatenated to provide one out-of-sample prediction for every observation included in the evaluation.

Model performance is evaluated through F1 score, Area Under Curve (AUC), and TSS, which are defined as
\[
\mathrm{F1}=\frac{2\mathrm{TP}}{2\mathrm{TP}+\mathrm{FP}+\mathrm{FN}},
\]
\[
\mathrm{AUC}=\int_0^1 \mathrm{TPR}(\mathrm{FPR})\,d(\mathrm{FPR}),
\]
where
\[
\mathrm{TPR}=\frac{\mathrm{TP}}{\mathrm{TP}+\mathrm{FN}}, \qquad
\mathrm{FPR}=\frac{\mathrm{FP}}{\mathrm{FP}+\mathrm{TN}},
\]
and
\[
\mathrm{TSS}=\mathrm{TPR}-\mathrm{FPR},
\]
given true positives (TP), false positives (FP), true negatives (TN), and false negatives (FN). To contextualize these scores under class imbalance, we also report a random baseline driven by fold prevalence. 
In fold $k$, let $q_k$ denote the proportion of training samples belonging to the positive (HSS) class. The random classifier then assigns each sample to the HSS class with probability $q_k$, and to the non-HSS class with probability $1-q_k$.
This yields an expected baseline with $\mathrm{AUC}=0.5$ and $\mathrm{TSS}=0$, while the other metrics depend on prevalence mismatch between train and test subsets.

Because the objective of this work is to investigate the predictive information contained in the topological representation rather than to optimize an operational forecasting system, the comparison across classifiers presented in Section \ref{sec:result_comparison} is primarily used to assess robustness of the information carried by the different topological vectorizations.

\subsection{Topological attribution and spatial evidence maps}\label{sec:xai}

The classification analysis described above establishes whether the topological representations contain information that can discriminate between images associated with HSS and non-HSS conditions. However, classification scores alone do not indicate which topological features drive the model decisions or where the corresponding structures are located in the original EUV observations. We therefore complement the classification analysis with an explainability procedure that connects the vectorized persistence features used by the classifier to their spatial localization in the original EUV image. A complete example illustrating all the steps of this procedure is presented in Section \ref{sec:working_example}.

\subsubsection{From model weights to topological relevance.}
For each model-vectorization pair, the cross-validation procedure produces $K$ independently trained models, one for each fold. Each model receives as input the $2d$ standardized topological features introduced above. After training, we extract from the model fitted in each fold a measure of the relevance assigned to each input feature. These $2d$ feature-relevance values describe, at the model level, which components of the topological representation are used to discriminate between the HSS and non-HSS classes. We have therefore the following weights
\begin{equation*}
w^{(k)}=(w^{(k)}[0],w^{(k)}[1])=\left(w^{(k)}_1,\ldots,w^{(k)}_{2d}\right),
\qquad k=1,\ldots,K,
\end{equation*}
where $K=5$ in the present analysis. For linear classifiers, $w^{(k)}_j$ corresponds directly to the coefficient associated with feature $j$. Its sign indicates the direction of the contribution to the decision function: positive coefficients are associated with the HSS class and negative coefficients with the non-HSS class. For tree-based models, $w^{(k)}_j$ is the model feature importance \cite{breiman2001,geurts2006,chen2016}; for GreedyMatern, it is permutation importance \cite{breiman2001,fisher2019}.

The resulting vector 
\begin{equation}
\bar{w} = \frac{1}{K}
\sum_{k=1}^{K}w^{(k)}
\label{eq:weights}
\end{equation}
provides a global, cross-validation-aggregated description of feature relevance. Features with large absolute values of $\bar{w}_j$ are those receiving a consistently strong contribution across folds, while, for signed relevance measures such as linear-model coefficients, the sign distinguishes contributions associated with the HSS and non-HSS classes.

\subsubsection{Projection onto the EUV image.}

The final step consists of relating the relevant topological features to structures in the original solar observation. Persistence features are defined in filtration space rather than directly in the spatial coordinates of the image. Their birth and death values, however, are associated with levels of the scalar field used to construct the filtration. This correspondence provides the link between the topological representation and the EUV image.

Due to its better performance with respect to the other vectorization strategies, which are reported in Table \ref{tab:top10}, in the following we restrict the methodology to the persistence image vectorization scheme. Consequently, with a slight abuse of notation, it is more comfortable to reshape both $x[h]^{\mathrm{std}}$ and $\bar{w}[h]$ from vectors in $\mathbb{R}^d$ to images in $\mathbb{R}^{\sqrt{d}}\times\mathbb{R}^{\sqrt{d}}$.

For a selected observation, we then build homology-specific contribution maps in persistence-image coordinates:
\[
c[h](i,j)=x[h]^{\mathrm{std}}(i,j)\,\bar{w}[h](i,j),
\qquad h\in\{0,1\},
\]
where $i$ indexes persistence bins and $j$ indexes birth bins. We then collapse along persistence bins to obtain birth-axis profiles
\[
s[h](j)=\sum_i C[h](i,j),
\qquad
s(j)=s[0](j)+s[1](j).
\]
Let $I_{\mathrm{proc}}(u,v)$ be the normalized/inverted scalar field used as the starting point for the TDA. If $N_h$ is the number of birth bins, we define the birth-bin index of each pixel as the \textit{quantization}
\[
j[h](u,v)=\left\lfloor \frac{I_{\mathrm{proc}}(u,v)-\min(I_{\mathrm{proc}}(u,v))}{\max(I_{\mathrm{proc}}(u,v))-\min(I_{\mathrm{proc}}(u,v))}\,(N_h-1)\right\rfloor,\qquad h\in\{0,1\},
\]
where $\lfloor\cdot\rfloor$ indicates the integer part or \textit{floor} function. This is a linear quantization of the pixel intensity into the same birth-axis discretization used by the persistence-image representation: each pixel is assigned to the birth-intensity band with which it is most naturally associated under the filtration (not to a unique topological generator). The projected fields are then obtained by:
\[
E[h](u,v)=s[h]\big(j[h](u,v)\big).
\]
In this way, each pixel receives the signed contribution value of its birth bin, and feature-level information is transformed into a spatial signed evidence field over the solar disk. Note that $E$ and $I_{\mathrm{proc}}$ are images of the same dimensions, which is crucial for the interpretation of the topological analysis on the original signal.


Positive evidence in $E[h]$ denotes image regions whose associated topological features support classification as HSS, whereas negative evidence denotes regions whose associated features support the non-HSS class.

Operationally, we first compute the signed projected field $E[h]$ on disk pixels, then smooth it with a Gaussian filter (default $\sigma=2$ pixels) to reduce discretization artifacts. 
To identify the regions displaying sufficiently strong
evidence, we define separate thresholds for the positive
and negative components
\[
\tau{[h]}_+
=
Q_{\xi}\left(
\{E{[h]}_+(u,v):(u,v)\in\Omega\}
\right),
\]
\[
\tau{[h]}_-
=
Q_{\xi}\left(
\{E{[h]}_-(u,v):(u,v)\in\Omega\}
\right),
\]
where $\Omega$ denotes the set of pixels belonging to
the solar disk and $Q_{\xi}$ is the empirical quantile
operator of order $\xi\in[0,1]$. In this work, we set
$\xi=0.55$, corresponding to the 55th percentile of
the respective evidence distributions.
Then we compute 
\begin{align*}
R[h](u,v)=\frac{\max(E[h]_+(u,v)-\tau[h]_+,0)}{\max(E[h]_+)-\tau[h]_+ + \varepsilon},\\
B[h](u,v)=\frac{\max(E[h]_-(u,v)-\tau[h]_-,0)}{\max(E[h]_-)-\tau[h]_- + \varepsilon},
\end{align*}
with small $\varepsilon>0$. This produces graded red (R) and blue (B) overlays on the EUV image instead of binary masks. Thanks to the thresholding produced by the quantile, pixels become active only when they exceed the corresponding onset, and are then displayed with continuous intensity proportional to exceedance.
Finally, we regularize the supports with morphology (opening/closing), keep connected components larger than a minimum area (default 120 pixels, with largest-component fallback), and apply one dilation step for visual continuity. These operations suppress isolated speckles and preserve coherent structures while retaining the sign and relative strength of the evidence.


The resulting maps should not be interpreted as segmentations of CHs, nor do they impose an a priori association between a given homology class and a particular solar structure. Rather, they provide a spatial representation of the regions of the EUV observation associated with topological features contributing to the classifier decision. This allows us to investigate, a posteriori, whether the predictive topological evidence is spatially associated with CHs or with other structures present in the solar corona.

\section{Results and Discussion}\label{section:results_discussion}

\subsection{Global performance of classification methods}\label{sec:result_comparison}
\begin{figure}[H]
\centering
\begin{subfigure}{0.9\textwidth}
\centering
\includegraphics[width=\textwidth]{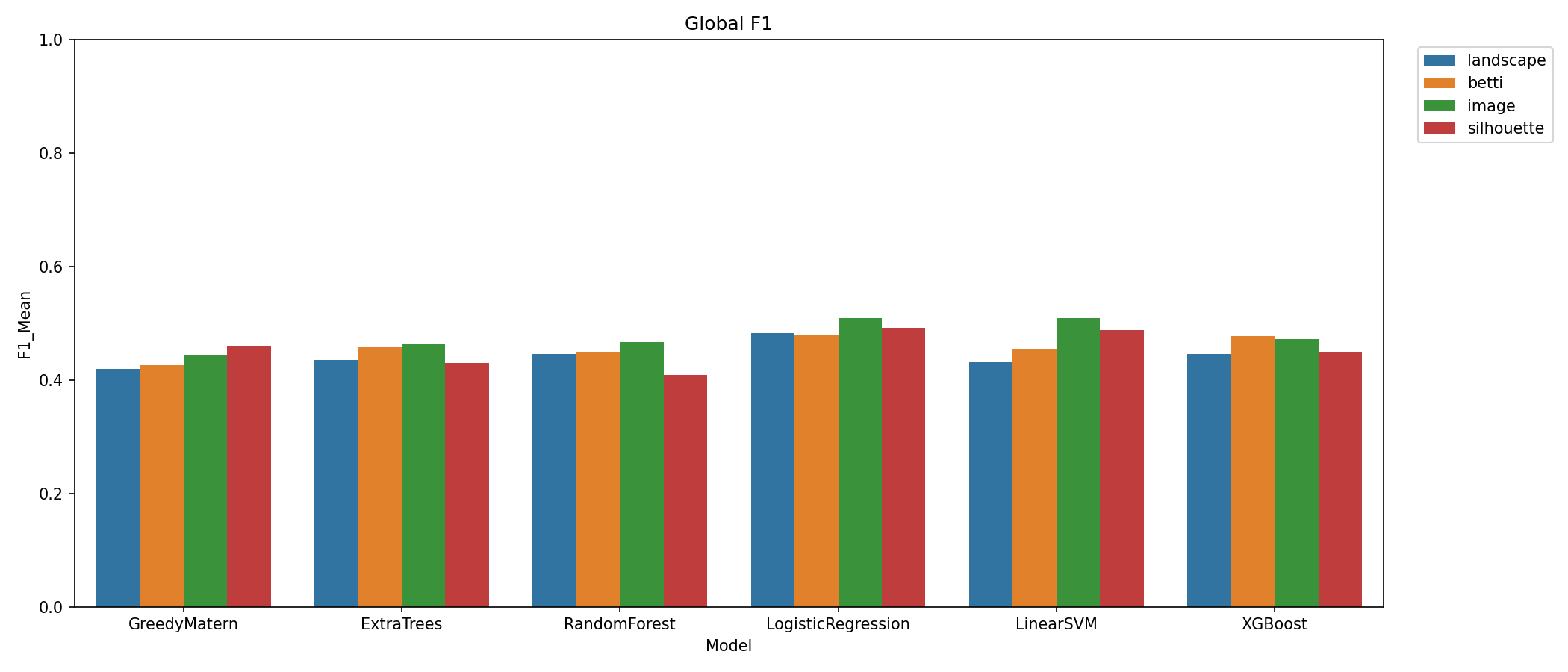}
\caption{F1.}
\end{subfigure}
\hfill
\begin{subfigure}{0.9\textwidth}
\centering
\includegraphics[width=\textwidth]{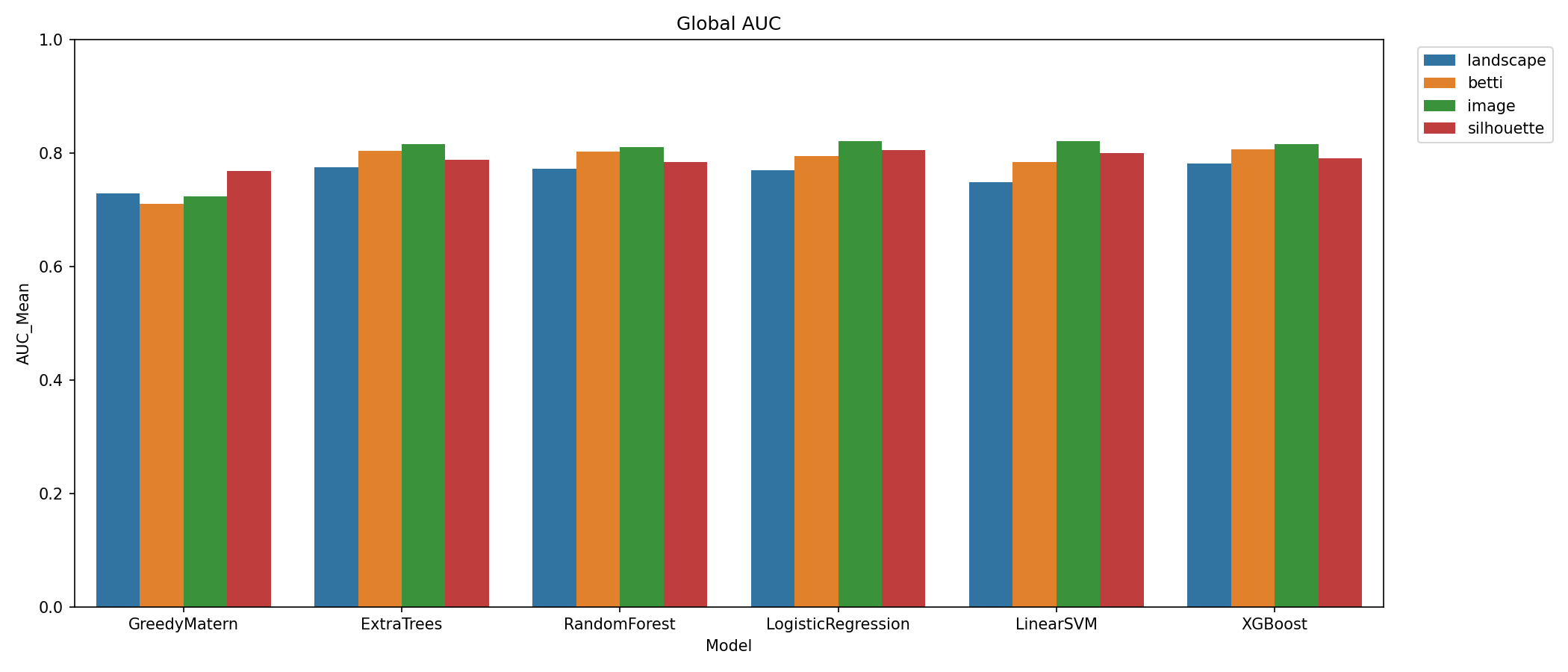}
\caption{AUC.}
\end{subfigure}
\hfill
\begin{subfigure}{0.9\textwidth}
\centering
\includegraphics[width=\textwidth]{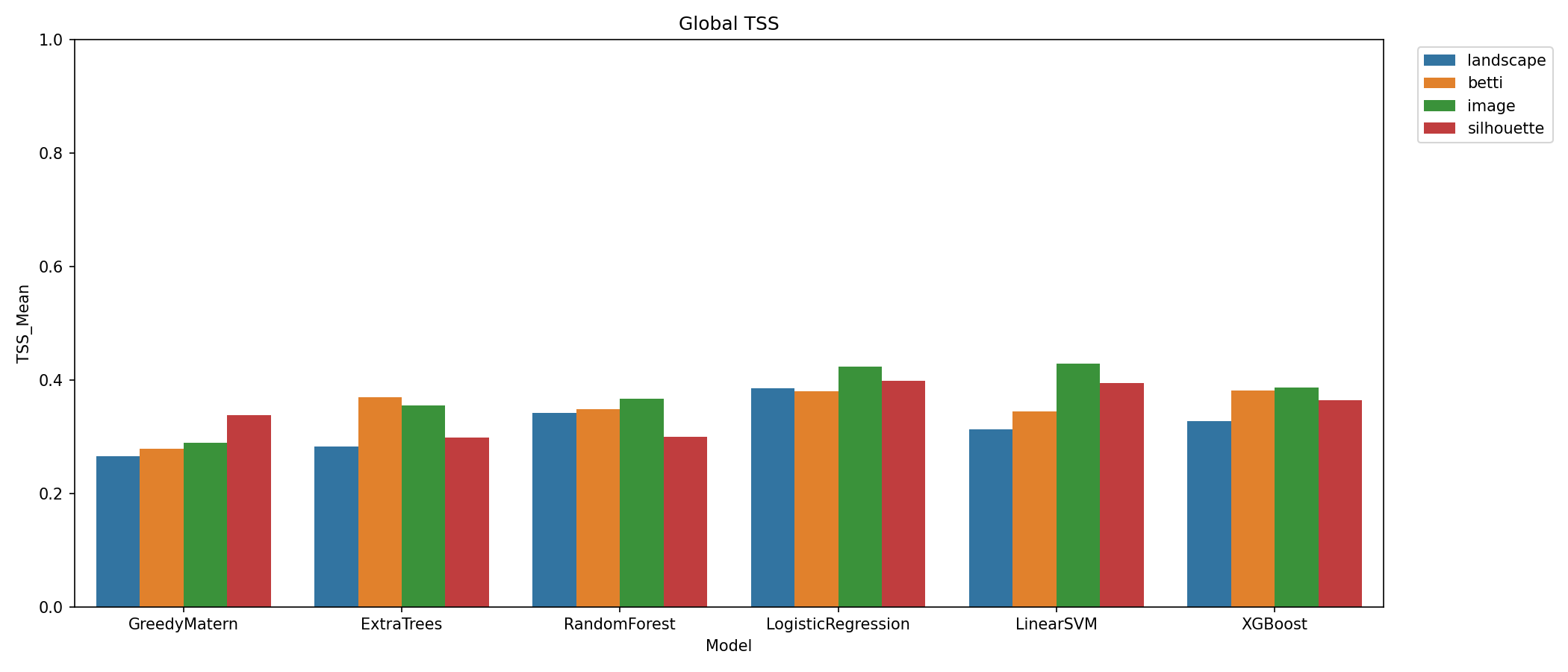}
\caption{TSS.}
\end{subfigure}
\caption{Performance comparison across the different
model--vectorization combinations for HSS prediction. The panels show the average
performance in terms of (a) F1 score, (b) AUC, and
(c) TSS.}
\label{fig:barre}
\end{figure}

In Table \ref{tab:top10}, we show the top 10 model/vectorization configurations according to the mean performance achieved in the test folds considered. Moreover, in Figure \ref{fig:barre} we show the global performance of all methods.

\begin{table}[h]
\centering
\caption{Performance comparison of the ten best-performing
model--vectorization combinations for HSS prediction, ranked by the F1 score.
Performance is evaluated in terms of F1 score,
AUC,
and TSS.
Bold values indicate the highest score achieved for each metric
among the reported combinations.}
\label{tab:top10}
\begin{tabular}{llccc}
\toprule
Model & Vectorization & F1 & AUC & TSS \\
\midrule
LogisticRegression & image & \textbf{0.5093} & \textbf{0.8204} & 0.4238 \\
LinearSVM & image & 0.5090 & 0.8203 & \textbf{0.4288} \\
LogisticRegression & silhouette & 0.4924 & 0.8057 & 0.3979 \\
LinearSVM & silhouette & 0.4875 & 0.8003 & 0.3937 \\
XGBoost & image & 0.4714 & 0.8156 & 0.3864 \\
LogisticRegression & landscape & 0.4829 & 0.7700 & 0.3848 \\
XGBoost & betti & 0.4771 & 0.8064 & 0.3816 \\
LogisticRegression & betti & 0.4786 & 0.7948 & 0.3794 \\
ExtraTrees & betti & 0.4581 & 0.8034 & 0.3692 \\
RandomForest & image & 0.4666 & 0.8110 & 0.3670 \\
\bottomrule
\end{tabular}
\end{table}

For the same folds and labels, the prevalence-driven random baseline is computed as follows. In each fold $k$, we estimate the prevalence of the HSS class in the training subset as follows
\[
q_k^{\mathrm{train}}=\frac{N_{\mathrm{HSS,train}}^{(k)}}{N_{\mathrm{train}}^{(k)}},
\]
where $N_{\mathrm{HSS,train}}^{(k)}$ denotes the number of class 1 and $N_{\mathrm{train}}^{(k)}$ the number of samples in the training subset of fold $k$, respectively, then, for each test sample in that fold, we draw a random prediction from $\mathrm{Bernoulli}(q_k^{\mathrm{train}})$. This uses only training information to set the random predictor and applies it on the test subset.

Let $q_k^{\mathrm{test}}$ be test prevalence in fold $k$. Under this baseline, expected fold metrics are
\[
\mathrm{F1}_k=\frac{2\,q_k^{\mathrm{test}}q_k^{\mathrm{train}}}{q_k^{\mathrm{test}}+q_k^{\mathrm{train}}},
\qquad
\mathrm{AUC}_k=0.5,
\qquad
\mathrm{TSS}_k=0.
\]
The resulting mean baseline is F1 $=0.1956$, AUC $=0.5$, and TSS $=0$, substantially lower than all reported TDA-based models in F1, AUC, and TSS. A complete overview for all folds is displayed in Table \ref{tab:baseline}.

\begin{table}[h]
\centering
\caption{Fold-wise class distribution and prevalence-driven random baseline. In fold $k$, random predictions use $q_k^{\mathrm{train}}$. The last column reports $\Delta\mathrm{F1}_k=\mathrm{F1}^{\mathrm{rand}}_k-\mathrm{F1}^{\mathrm{LogReg+image}}_k$. The second and third columns represent the number of positive samples and total number of samples in the test set associated to $k$ fold.}
\label{tab:random_baseline_folds}
\begin{tabular}{cccccc}
\toprule
Fold & $\#$ Test HSS & $\#$ Test non-HSS & $q_k^{\mathrm{train}}$ & F1 & $\Delta$F1 vs LogReg+image \\
\midrule
0 & 2496 & 10704 & 0.1937 & 0.1913 & -0.2665 \\
1 & 2928 & 11016 & 0.2008 & 0.2053 & -0.1218 \\
2 & 2592 & 11448 & 0.1983 & 0.1912 & -0.3286 \\
3 & 3216 & 12480 & 0.1888 & 0.1965 & -0.3924 \\
4 & 2760 & 12336 & 0.2054 & 0.1935 & -0.3276 \\
\bottomrule
\end{tabular}\label{tab:baseline}
\end{table}

\begin{figure}[h]
\centering
\begin{subfigure}{0.46\textwidth}
\centering
\includegraphics[width=\textwidth]{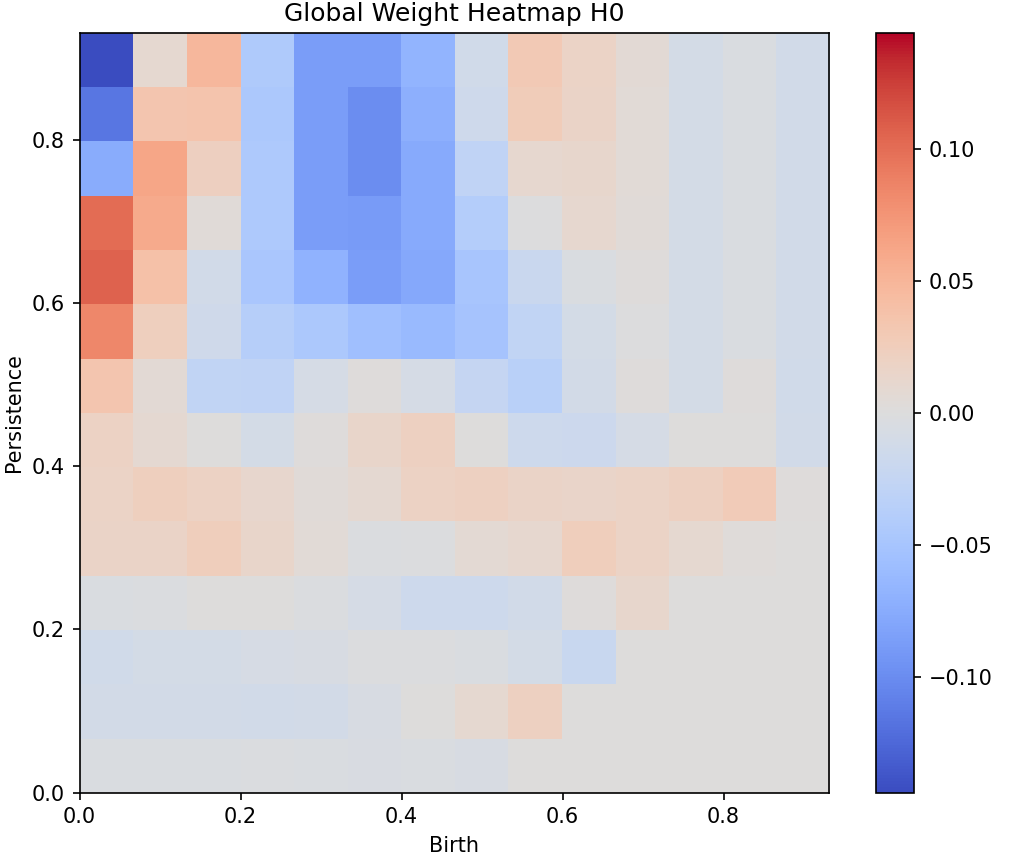}
\caption{$H_0$ global weights.}
\end{subfigure}
\begin{subfigure}{0.46\textwidth}
\centering
\includegraphics[width=\textwidth]{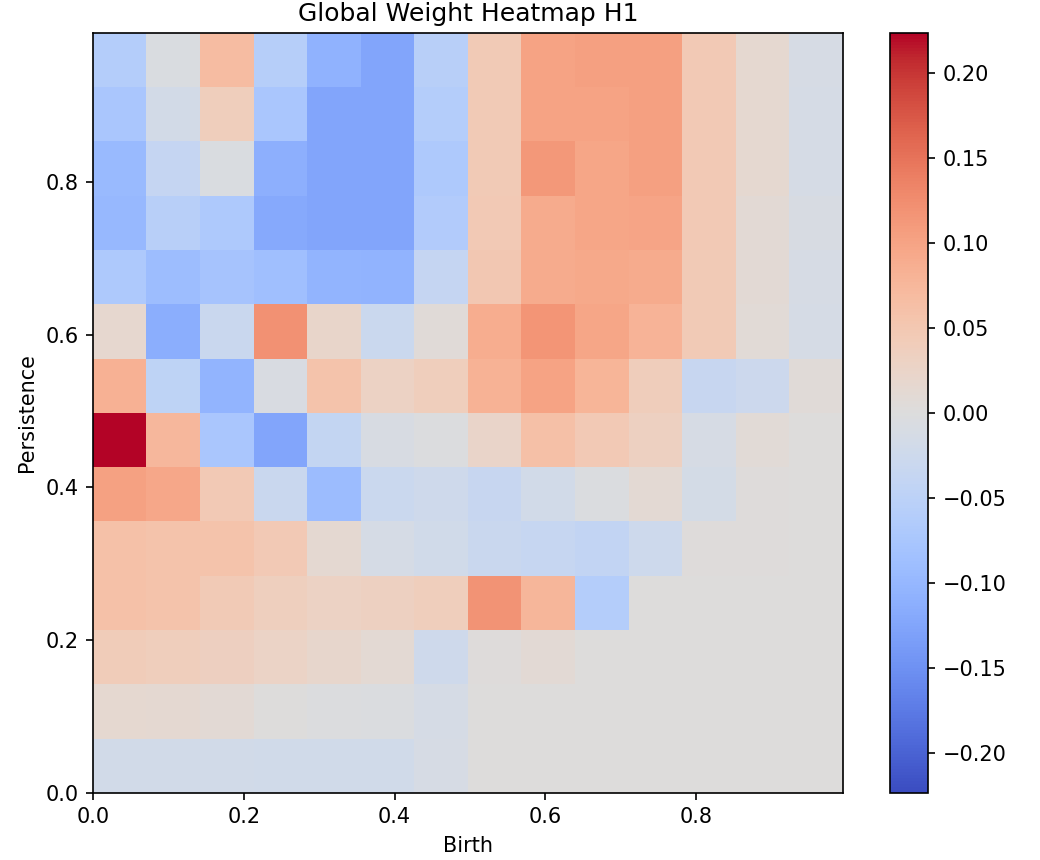}
\caption{$H_1$ global weights.}
\end{subfigure}
\caption{Global-weight explanation outputs for LogisticRegression+image, shown separately for $H_0$ and $H_1$ to improve readability of axes and colorbar labels.}
\label{fig:heatmap}
\end{figure}

The results reported show that both linear and non-linear approaches provide comparable performance, with no vectorization method consistently outperforming the others. This may suggest that the underlying data might not contain substantially more information that can be exploited to further improve performance.
Although \textit{LinearSVM+image} is first by TSS, \textit{LogisticRegression+image} is nearly tied and provides cleaner signed linear explainability, so we adopt it as primary interpretation model for our subsequent XAI. Therefore, in Figure \ref{fig:heatmap} we display the global weights (see equation \eqref{eq:weights}) corresponding to the image vectorization strategy coupled with the trained logistic classifier, both for $H_0$ and $H_1$ extracted features. Note that the dimension of the persistence image is $14\times 14$, which turned out to be a satisfactory trade-off between data representation and model complexity in our tests. The mixed-sign profile confirms that both HSS-supporting and quiet-supporting regions exist in feature space.

\subsection{Worked example: full chain from diagram to decision}\label{sec:working_example}

To clarify our XAI methodology, in the following we track a correctly classified positive sample through the entire pipeline.

\begin{figure}[h]
\centering
\includegraphics[width=0.32\textwidth]{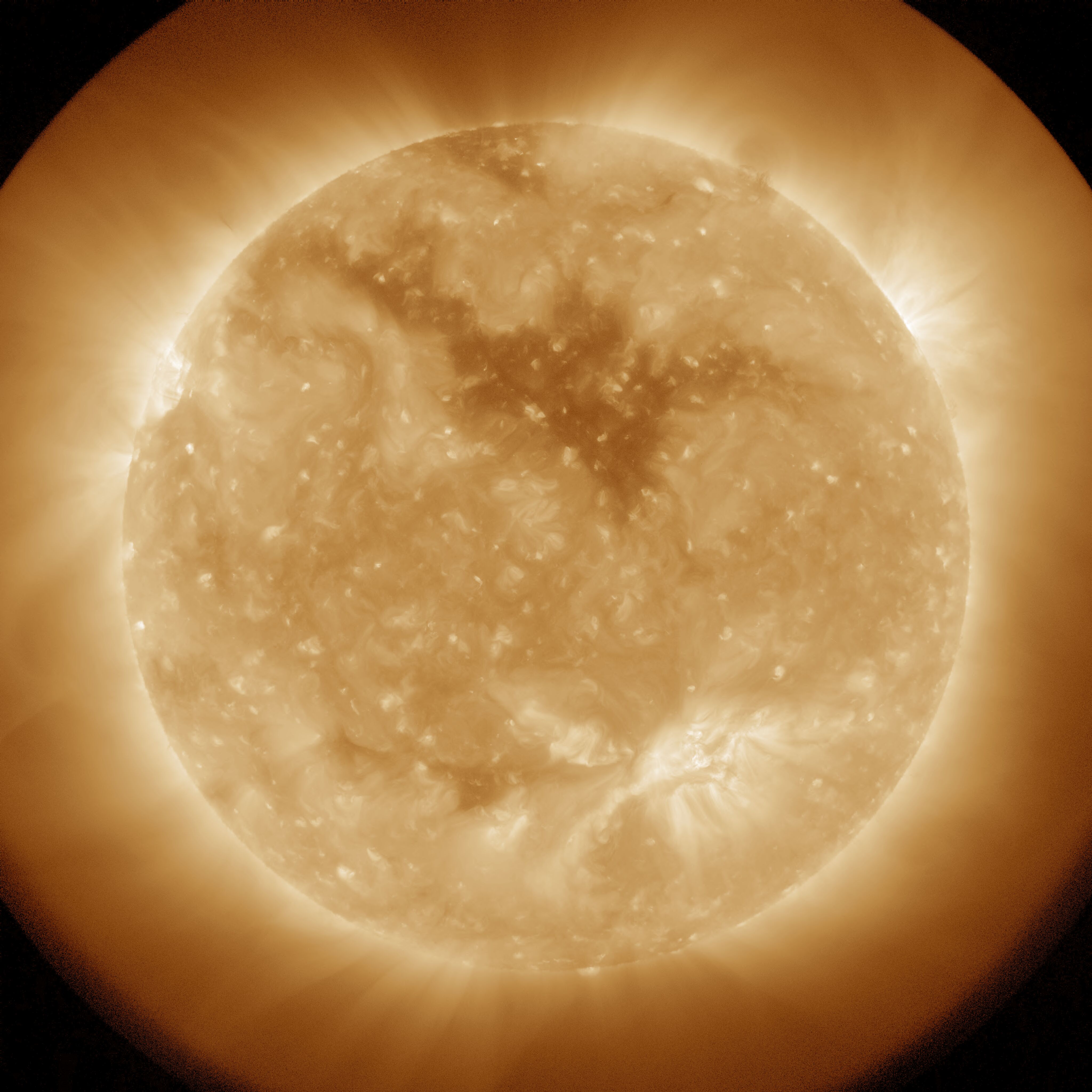}
\includegraphics[width=0.55\textwidth]{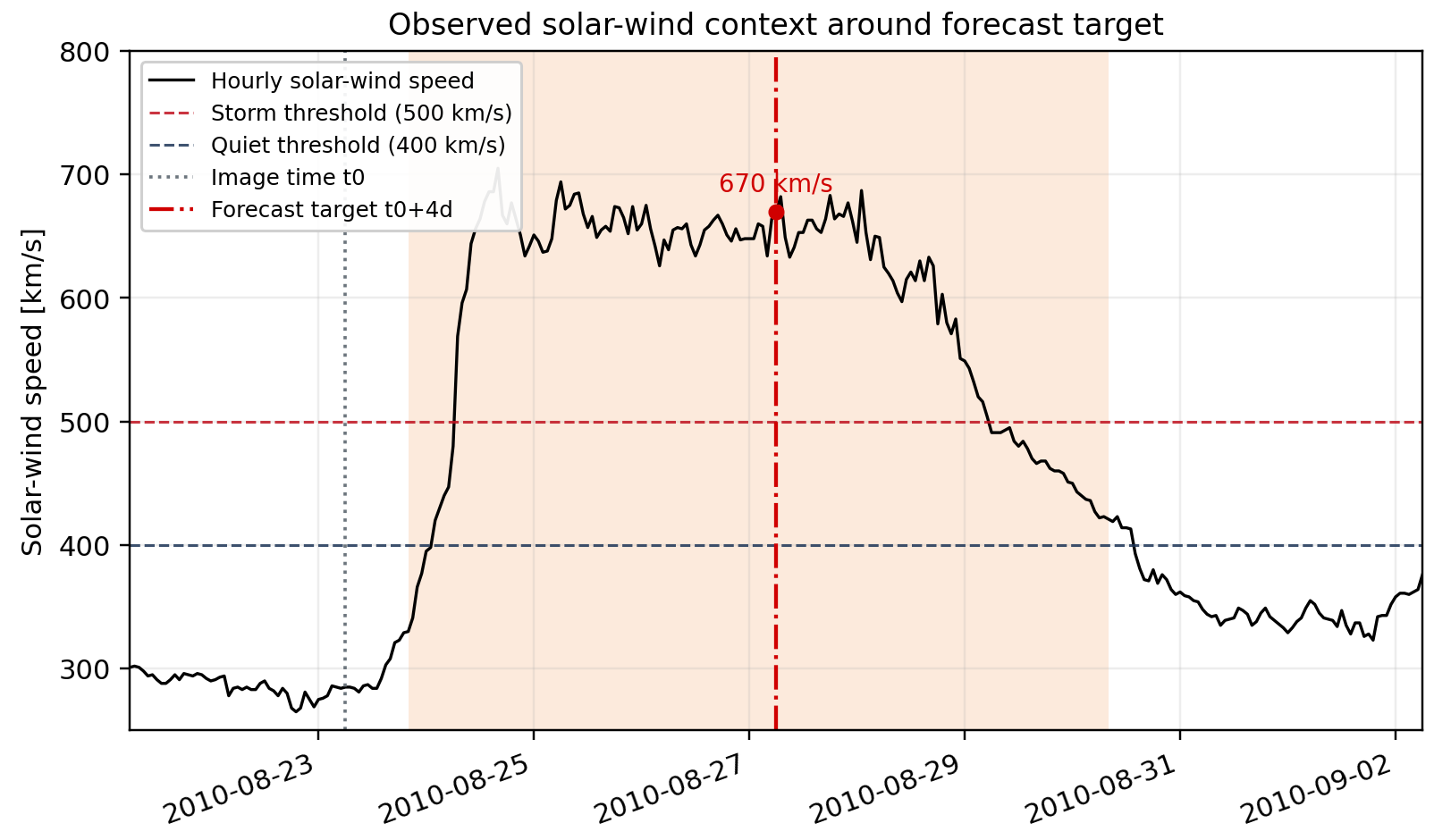}
\caption{Worked-example context. Left: original AIA 193~\AA\ image at $t_0=$ 2010-08-23 06:00. Right: hourly solar wind speed around the corresponding forecast window, with non-HSS/HSS thresholds and temporal intervals used for labeling.}
\label{fig:context_example}
\end{figure}

Before discussing the topological pipeline itself, it is useful to place this sample in its temporal forecasting context. The left panel of Figure \ref{fig:context_example} shows the original AIA 193~\AA\ image at time $t_0=$ 2010-08-23 06:00, while the right panel reports the hourly solar wind speed time series around the forecast target time $t_0+4$ days. The plotted thresholds ($400$ and $500$ km/s) are the same operational cutoffs used for non-HSS/HSS labeling in this work, and highlighted HSS intervals come from the considered catalog. This links the image-level morphology to the downstream event definition before entering the persistent-homology steps.

\subsubsection*{Step 1: persistence diagram}

\begin{figure}[h]
\centering
\begin{subfigure}{0.48\textwidth}
\centering
\includegraphics[width=\textwidth]{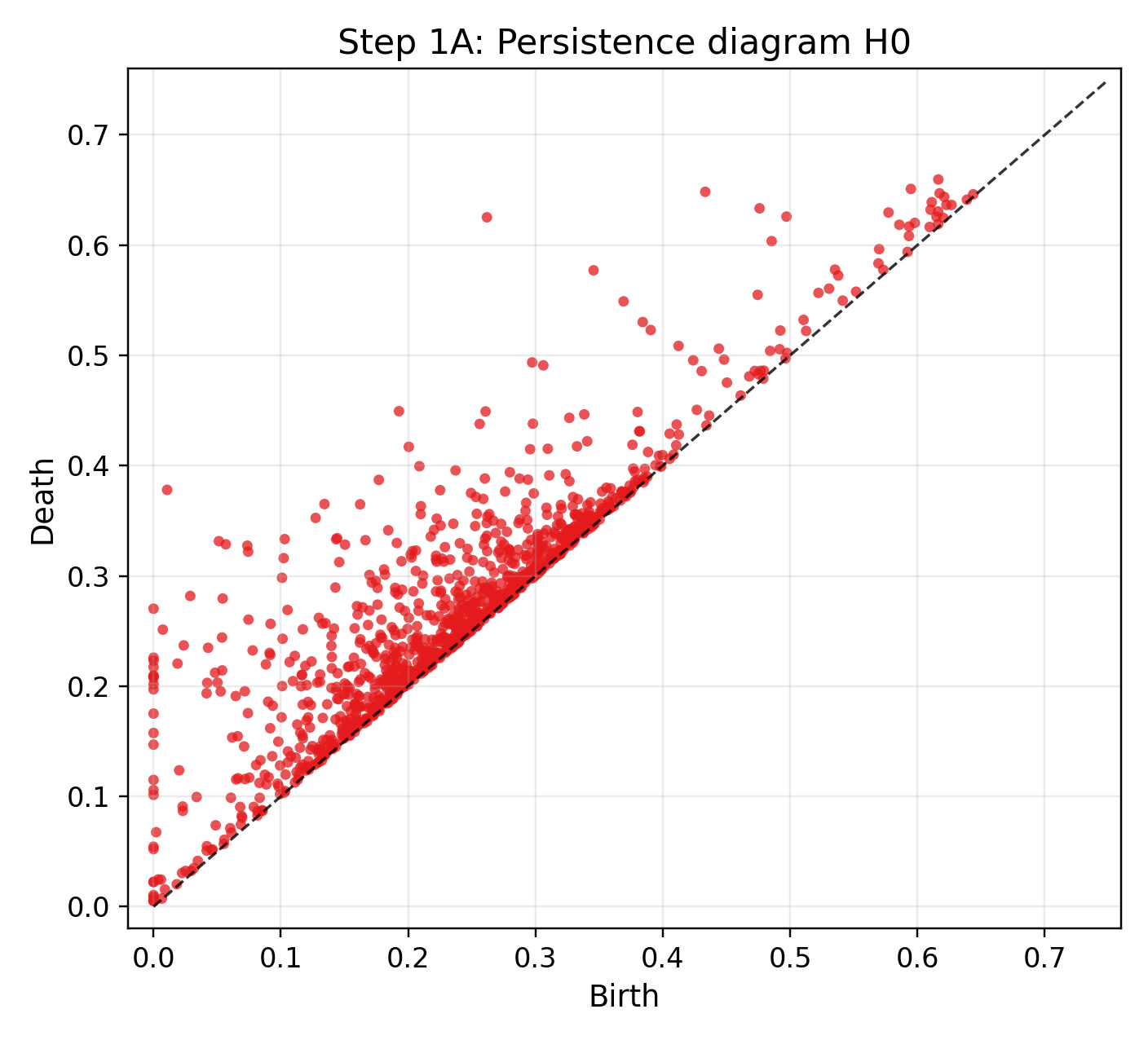}
\caption{$H_0$ persistence diagram.}
\end{subfigure}
\hfill
\begin{subfigure}{0.48\textwidth}
\centering
\includegraphics[width=\textwidth]{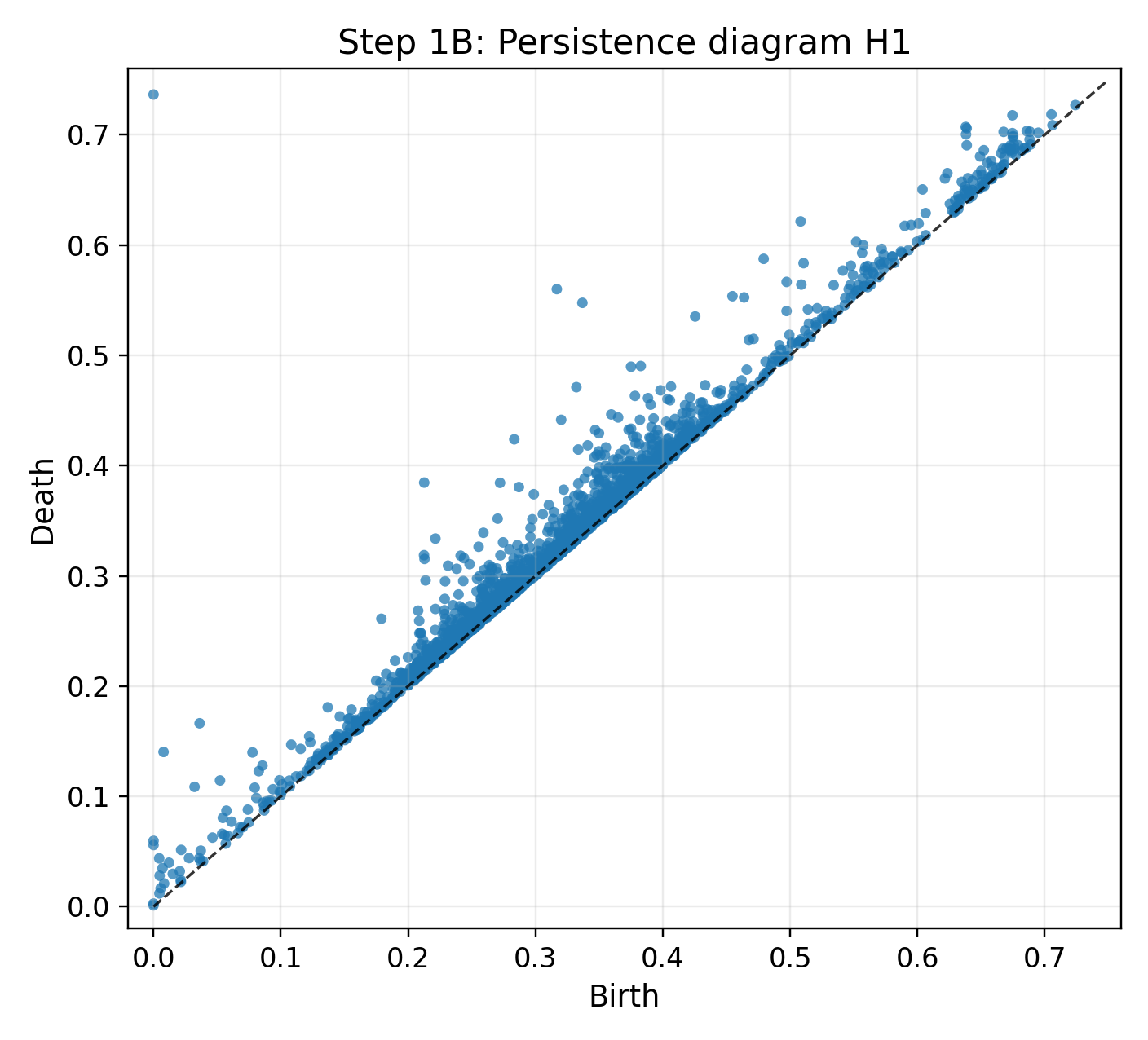}
\caption{$H_1$ persistence diagram.}
\end{subfigure}
\caption{Persistence diagrams of the selected sample shown separately for $H_0$ and $H_1$ (left and right panels, respectively). The maximum finite birth/death coordinate is $0.7366$ (max $H_1$ death), so there are no finite generators above that value and the axis cap at $0.75$ is data-driven.}
\label{fig:pd_example}
\end{figure}

The first step consists of processing the original AIA image (normalization and inversion) and then computing the persistence diagram for both $H_0$ and $H_1$ homologies out of the processed image (Figure \ref{fig:pd_example}). In light of the MNIST examples in Section \ref{sec:homology}, the contrast is evident: instead of one or two clearly isolated dominant loops (as for the digits ``0'' and ``8''), the solar diagram contains a much denser population of generators across scales.
This means that the coronal morphology around this event is topologically richer than the synthetic handwritten-digit cases. After normalization and inversion, CH-related boundaries and other coronal structures produce persistent loop classes ($H_1$) that appear farther from the diagonal, while the many points close to the diagonal correspond to short-lived, fine-scale fluctuations. In other words, the solar case combines robust mesoscale structures with a broad background of transient small-scale topology.

\subsubsection*{Step 2: persistence-image vectorization}

\begin{figure}[h]
\centering
\begin{subfigure}{0.48\textwidth}
\centering
\includegraphics[width=\textwidth]{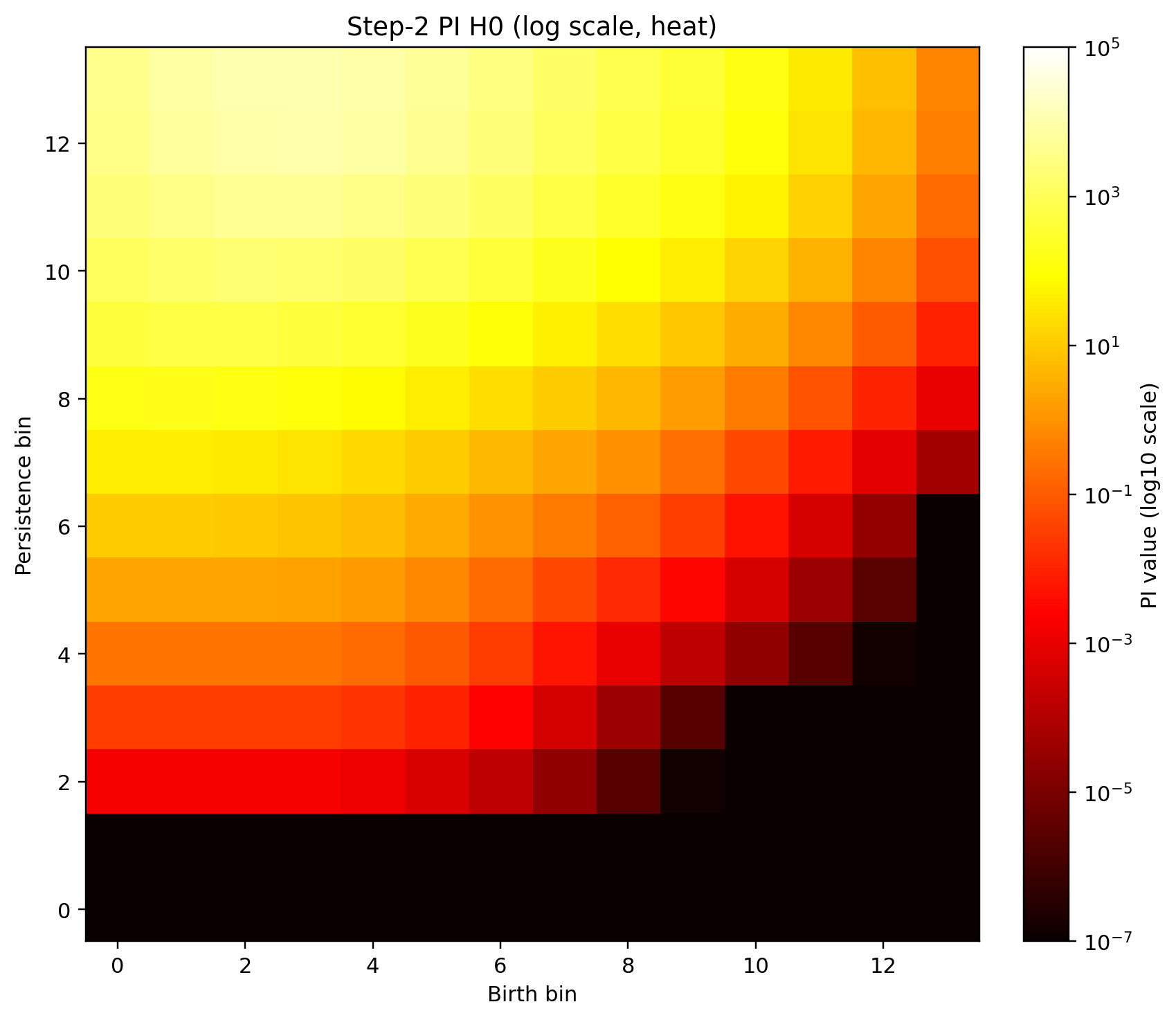}
\caption{$H_0$ persistence image.}
\end{subfigure}
\hfill
\begin{subfigure}{0.48\textwidth}
\centering
\includegraphics[width=\textwidth]{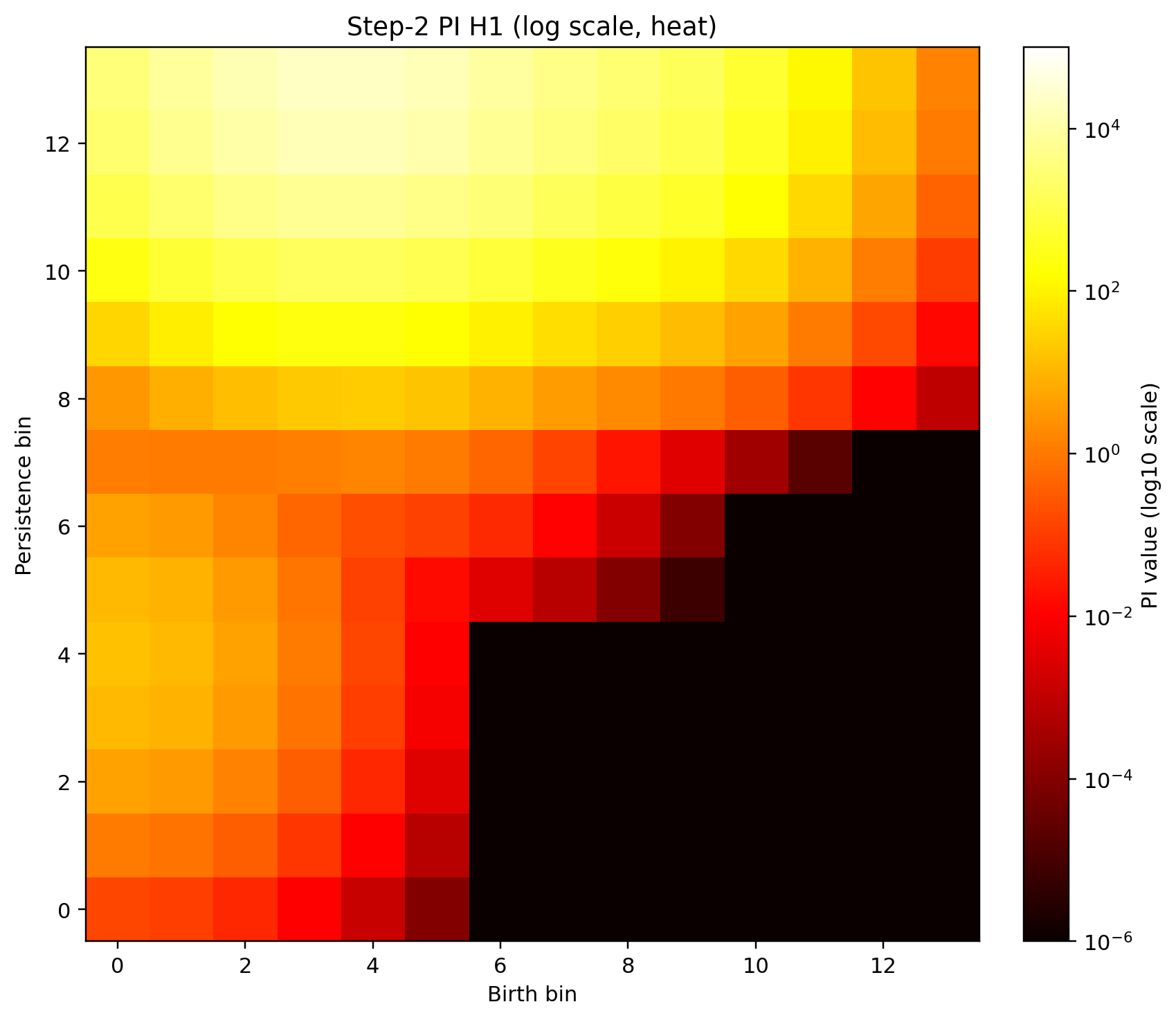}
\caption{$H_1$ persistence image.}
\end{subfigure}
\caption{Persistence-image vectorization ($14\times14$) shown separately for $H_0$ and $H_1$ (left and right panels, respectively). Image values are in logarithmic scale. Bins having zero values are colored in black.} 
\label{fig:images_example}
\end{figure}

The second step consists of vectorizing the persistence diagrams separately. This leads to two persistence images (Figure \ref{fig:images_example}). The visually sparse appearance is expected: most generators are concentrated at low persistence and around specific birth ranges, while bins receiving contributions from near-diagonal generators remain non-zero but comparatively small. Because color scaling is linear and dominated by the densest bins, these low-amplitude bins appear close to black although they are not exactly zero.

\subsubsection*{Step 3: feature contribution maps in persistence-image space}

\begin{figure}[h]
\centering
\begin{subfigure}{0.32\textwidth}
\centering
\includegraphics[width=\textwidth]{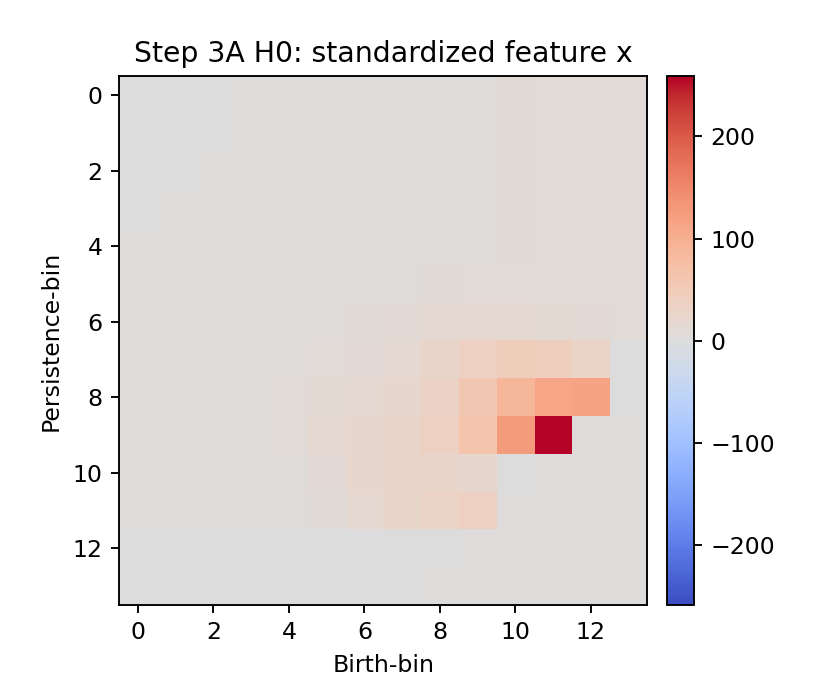}
\caption{$H_0$: $x_0^{\mathrm{std}}$.}
\end{subfigure}
\hfill
\begin{subfigure}{0.32\textwidth}
\centering
\includegraphics[width=\textwidth]{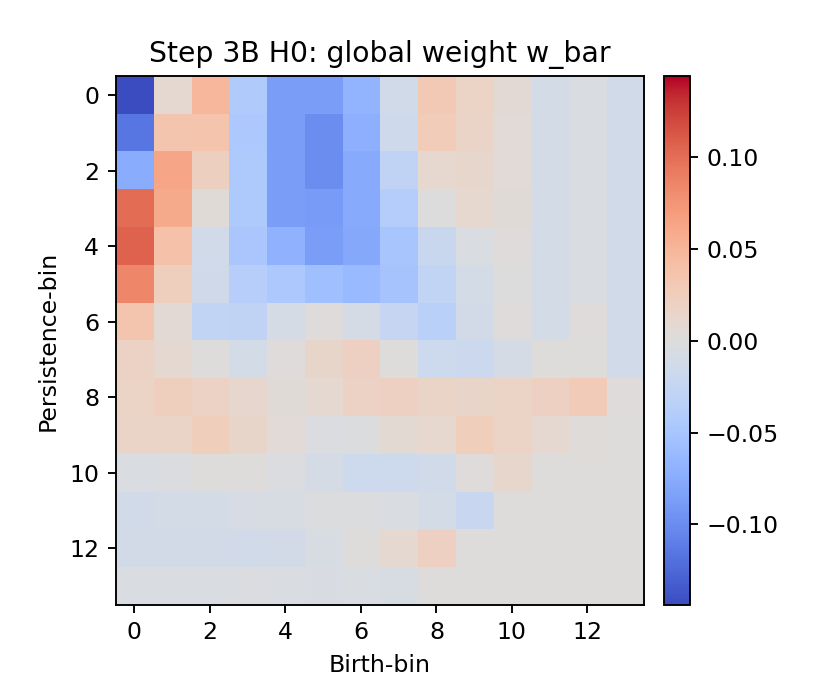}
\caption{$H_0$: $\bar{w}_0$.}
\end{subfigure}
\hfill
\begin{subfigure}{0.32\textwidth}
\centering
\includegraphics[width=\textwidth]{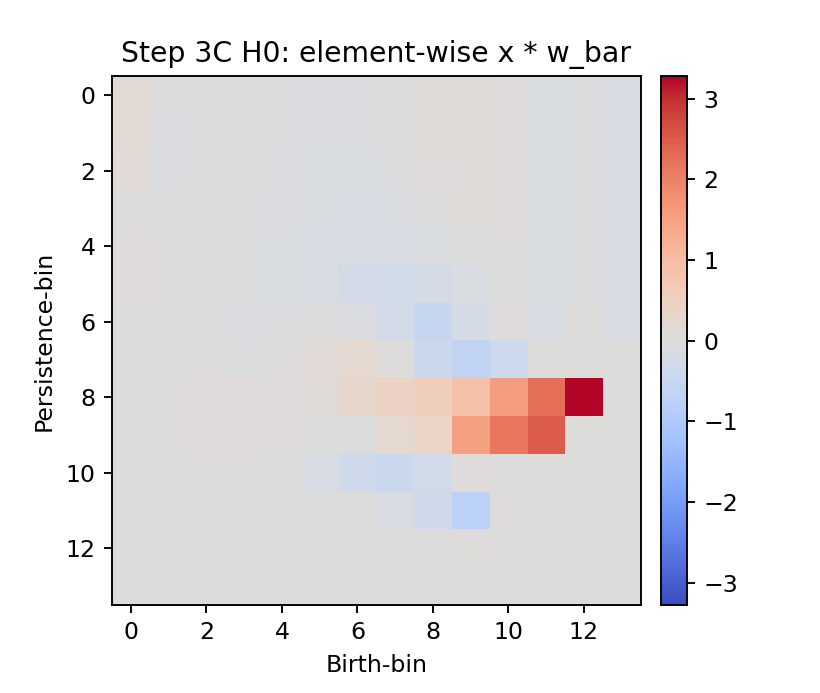}
\caption{$H_0$: $x_0^{\mathrm{std}}\odot\bar{w}_0$.}
\end{subfigure}

\begin{subfigure}{0.32\textwidth}
\centering
\includegraphics[width=\textwidth]{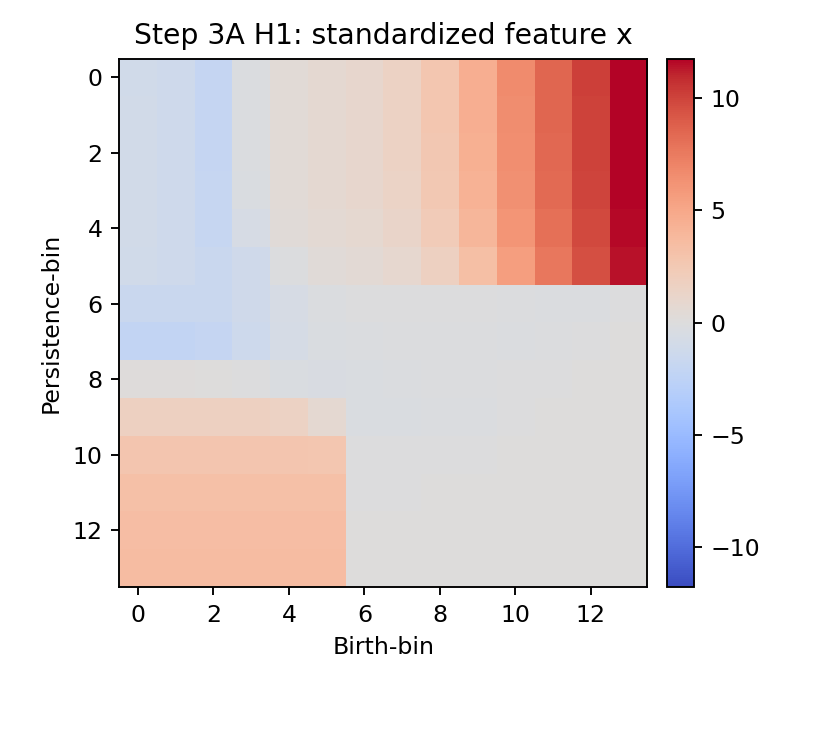}
\caption{$H_1$: $x_1^{\mathrm{std}}$.}
\end{subfigure}
\hfill
\begin{subfigure}{0.32\textwidth}
\centering
\includegraphics[width=\textwidth]{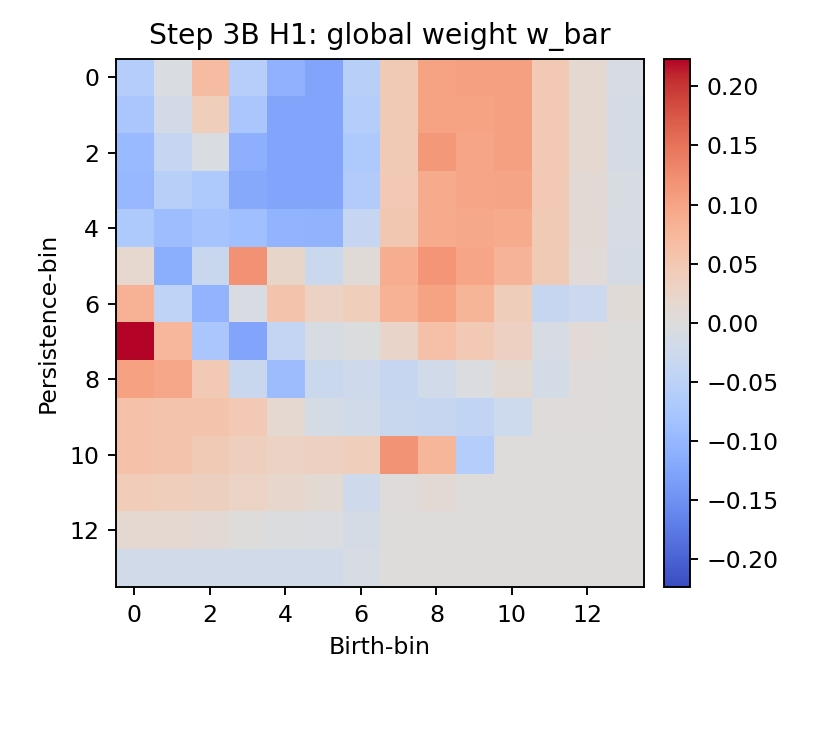}
\caption{$H_1$: $\bar{w}_1$.}
\end{subfigure}
\hfill
\begin{subfigure}{0.32\textwidth}
\centering
\includegraphics[width=\textwidth]{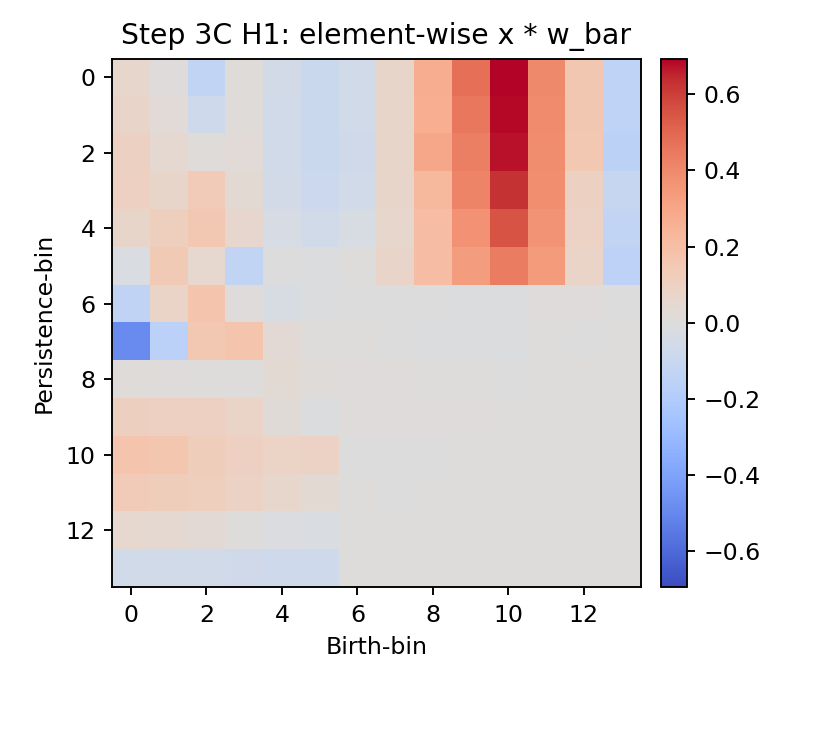}
\caption{$H_1$: $x_1^{\mathrm{std}}\odot\bar{w}_1$.}
\end{subfigure}
\caption{Step-3 decomposition with single-panel assets: standardized feature maps, global weight maps, and element-wise contributions for $H_0$ and $H_1$.}
\label{fig:multiply_example}
\end{figure}

Then, images are standardized and, according to the logistic regression model, they are then multiplied times the vector of weights as shown in Figure \ref{fig:multiply_example}.

\begin{figure}[h]
\centering
\begin{subfigure}{0.32\textwidth}
\centering
\includegraphics[width=\textwidth]{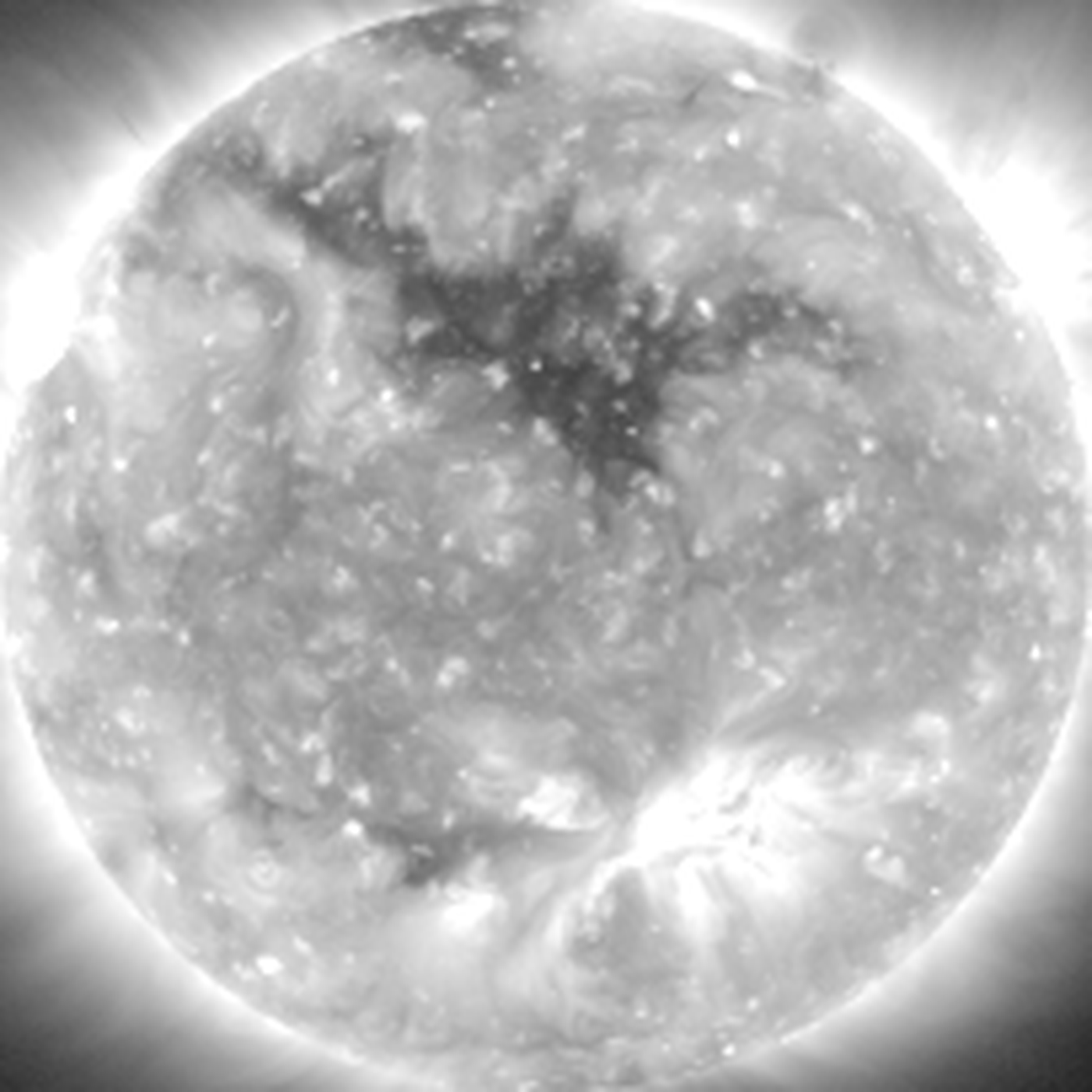}
\caption{Processed AIA.}
\end{subfigure}
\hfill
\begin{subfigure}{0.32\textwidth}
\centering
\includegraphics[width=\textwidth]{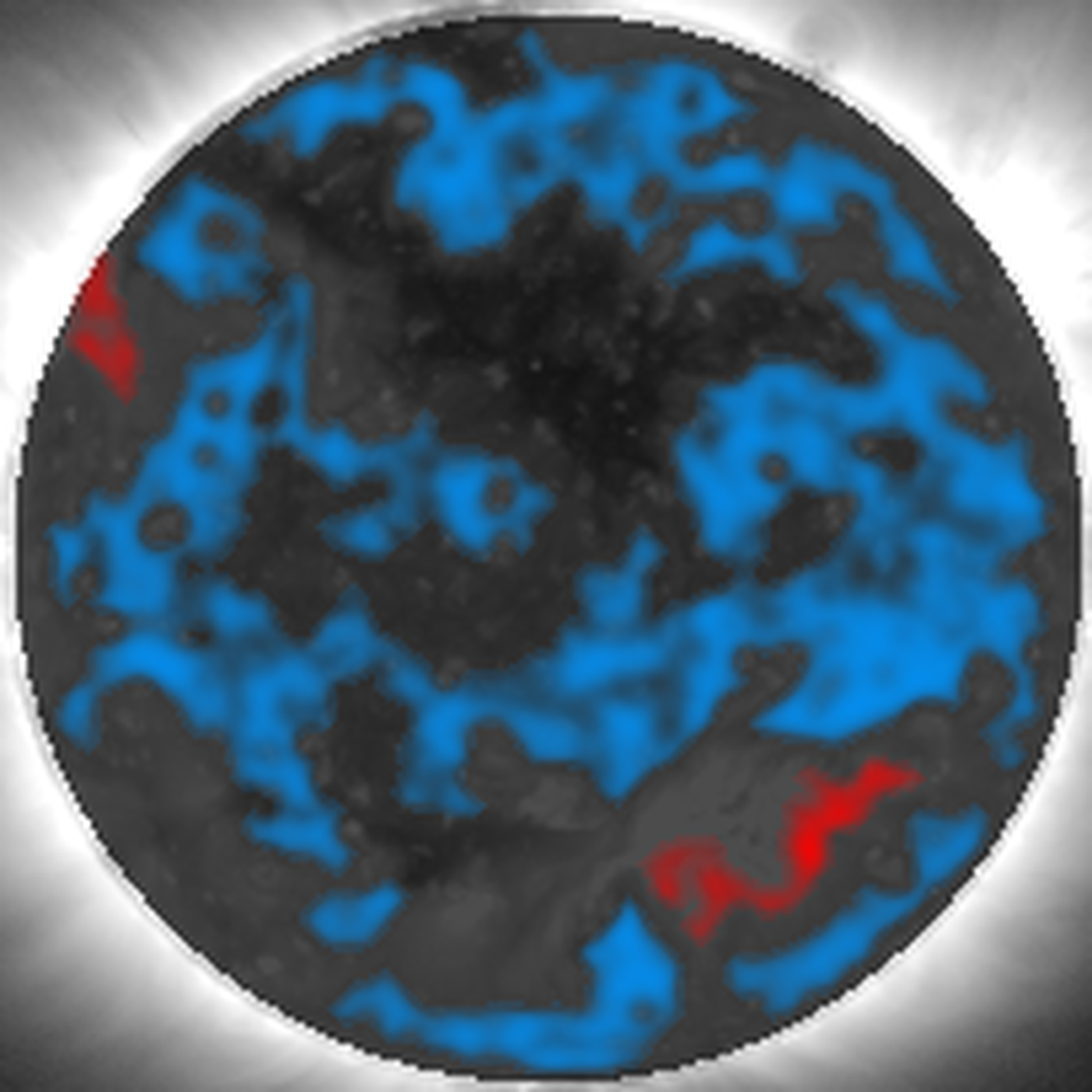}
\caption{$H_0$ evidence.}
\end{subfigure}
\hfill
\begin{subfigure}{0.32\textwidth}
\centering
\includegraphics[width=\textwidth]{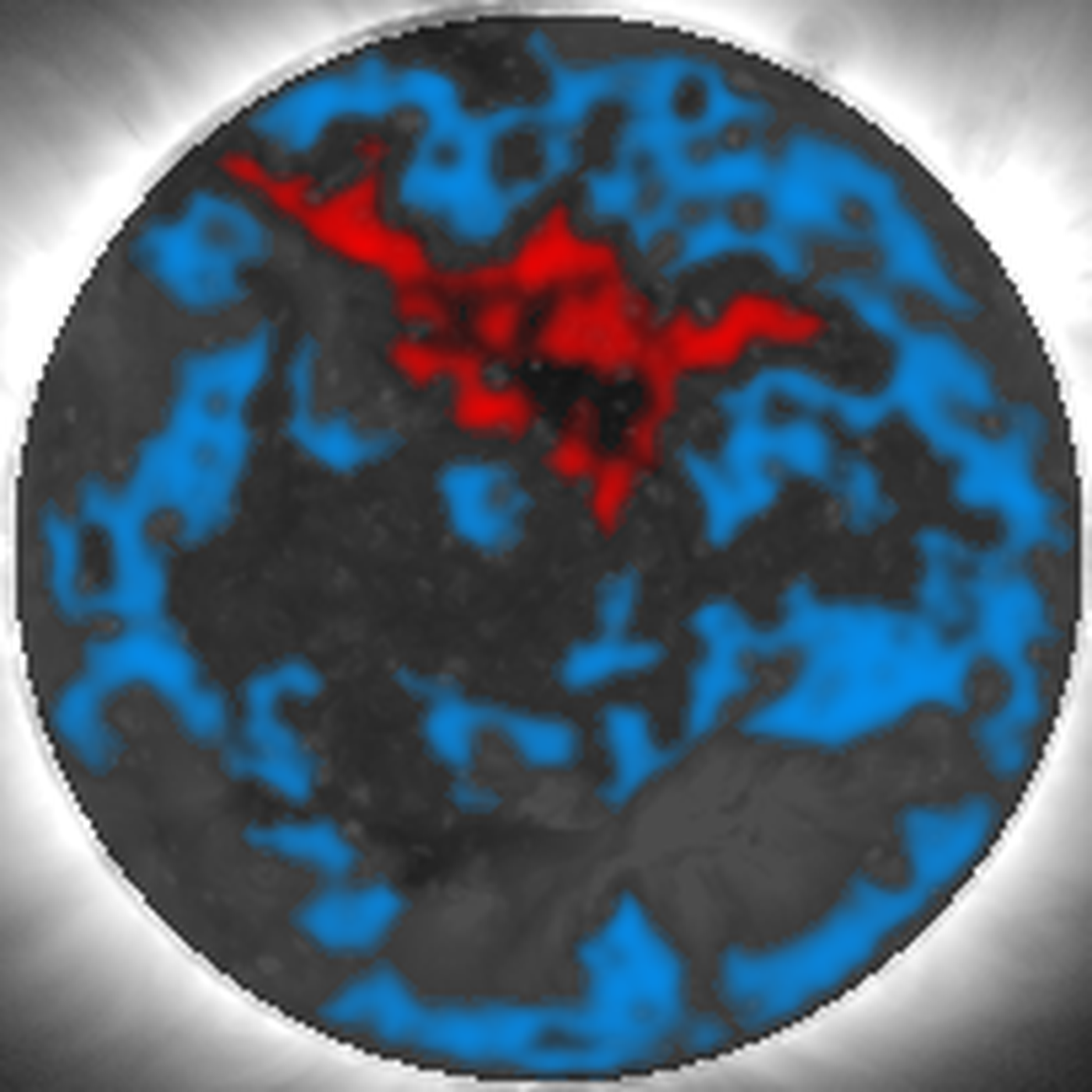}
\caption{$H_1$ evidence.}
\end{subfigure}
\caption{HSS example (2010-08-23 06:00): preprocessed AIA view with corresponding $H_0$ and $H_1$ projected evidence maps.}
\label{fig:storm}
\end{figure}

The transformation from persistence-image (PI) contributions to annotated solar images follows the procedure defined in Section \ref{sec:xai}: construction of\\ $C[h](i,j)$, birth-axis aggregation $s[h](j)$, projection through $j[h](u,v)$, and\\ quantile-thresholded continuous red/blue overlays after smoothing. Thus, Figure \ref{fig:multiply_example} is the PI-space decomposition, while Figure \ref{fig:storm} is its deterministic image-space projections.

\subsection{Analysis of topologically annotated samples}\label{sec:analysis_other}

\begin{figure}[h]
\centering
\begin{subfigure}{0.32\textwidth}
\centering
\includegraphics[width=\textwidth]{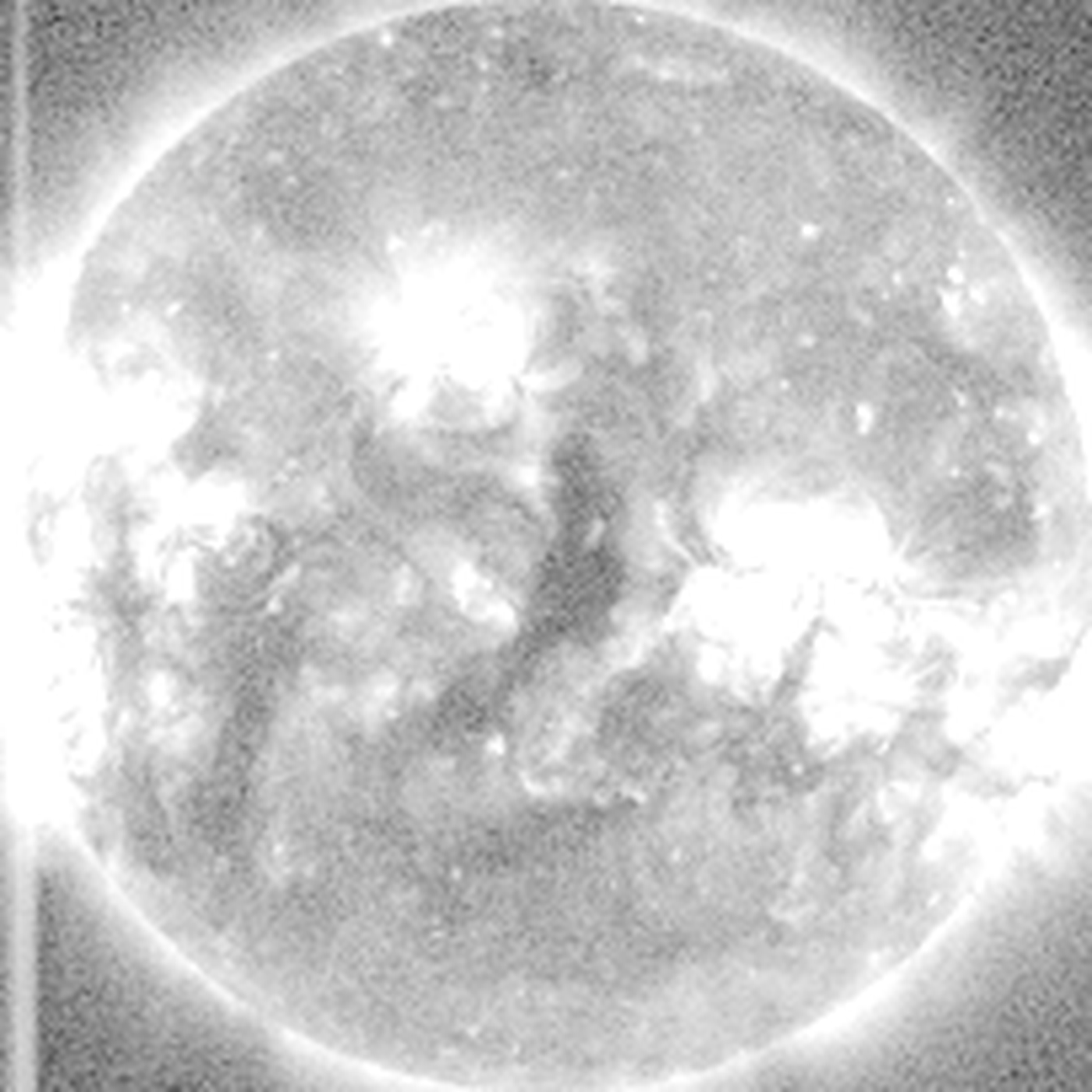}
\caption{Processed AIA.}
\end{subfigure}
\hfill
\begin{subfigure}{0.32\textwidth}
\centering
\includegraphics[width=\textwidth]{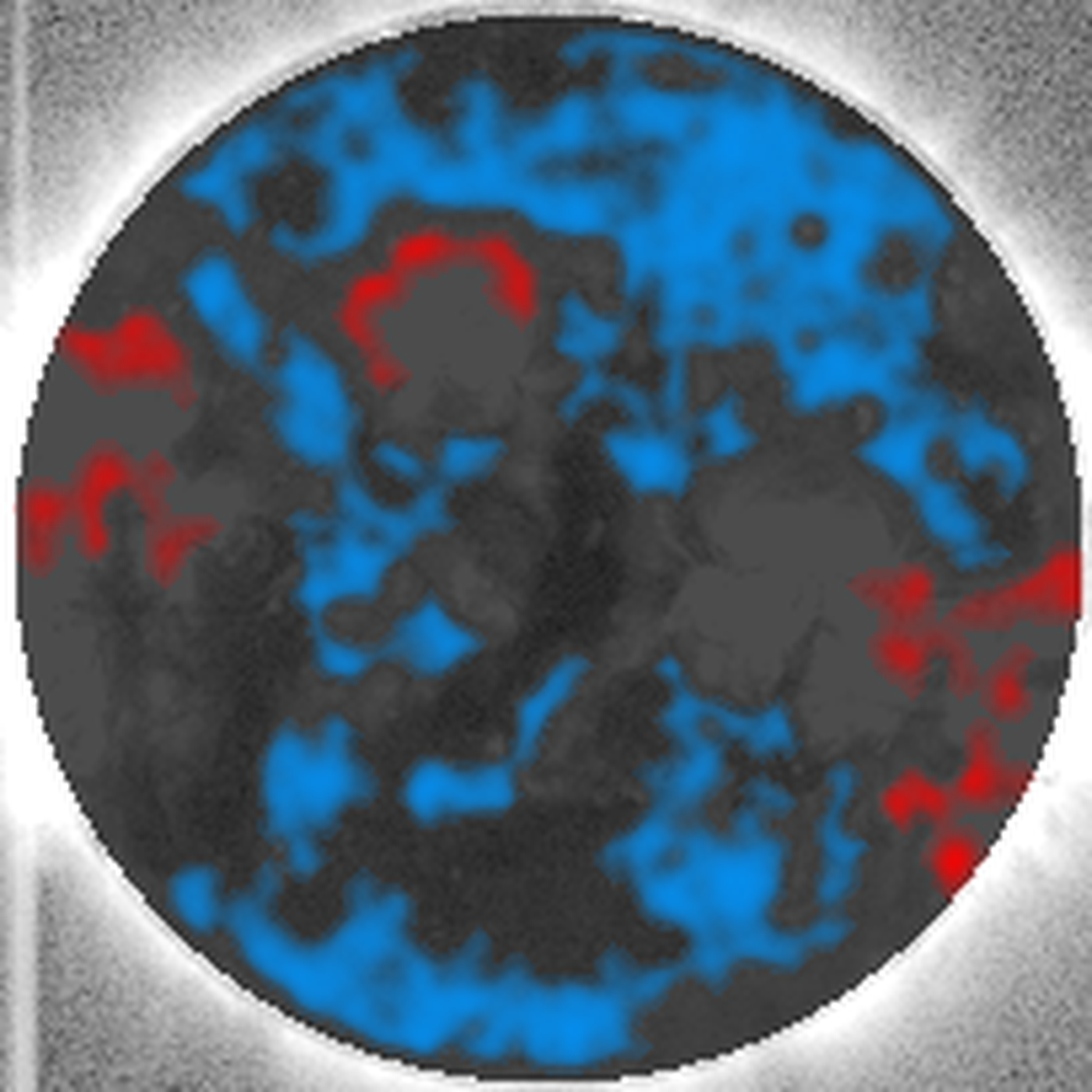}
\caption{$H_0$ evidence.}
\end{subfigure}
\hfill
\begin{subfigure}{0.32\textwidth}
\centering
\includegraphics[width=\textwidth]{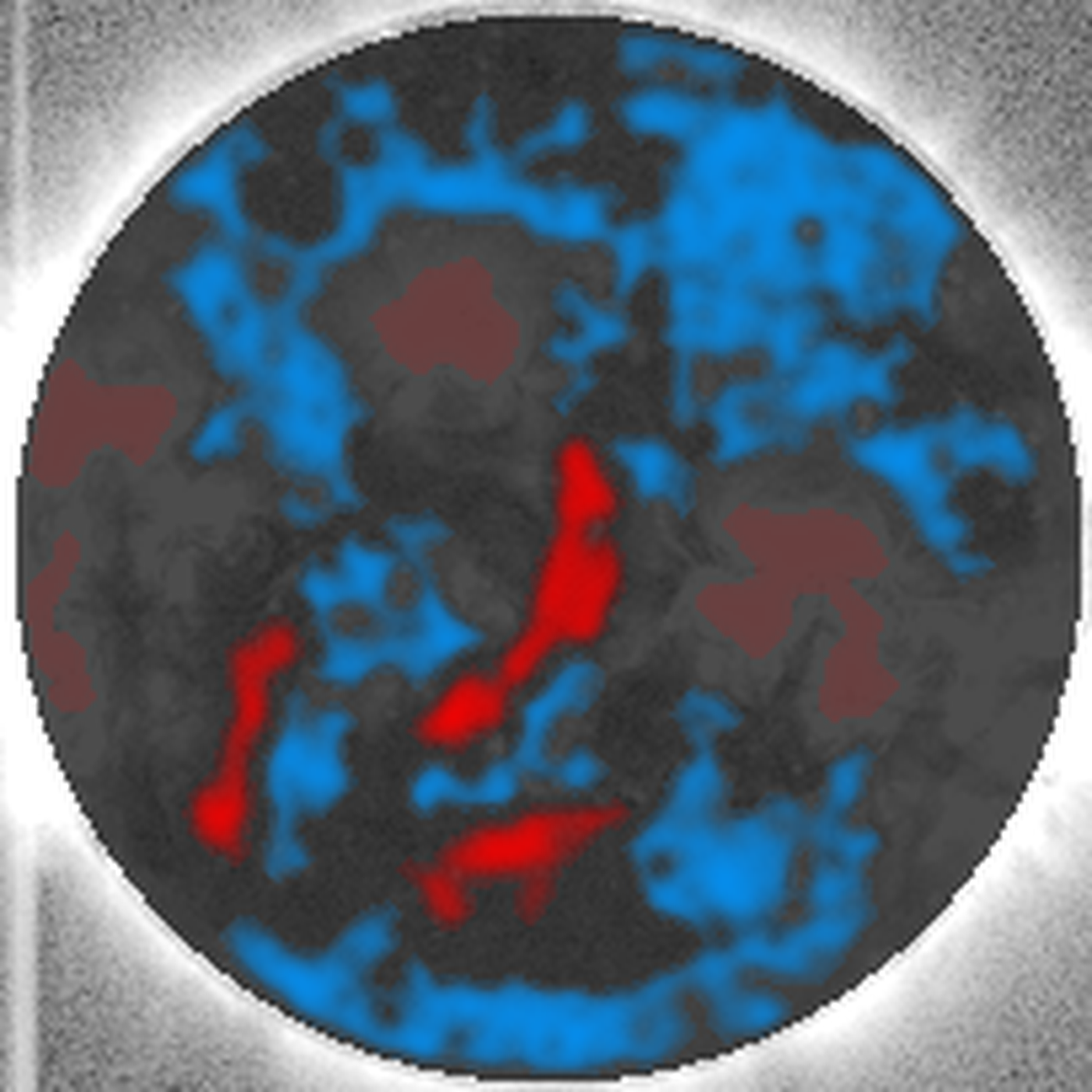}
\caption{$H_1$ evidence.}
\end{subfigure}
\caption{Non-HSS example (2014-02-25 01:00): processed AIA view with corresponding $H_0$ and $H_1$ projected evidence maps. Red indicates HSS-supporting projected evidence, blue indicates quiet-supporting projected evidence.}
\label{fig:quiet}
\end{figure}

\begin{figure}[h]
\centering
\begin{subfigure}{0.32\textwidth}
\centering
\includegraphics[width=\textwidth]{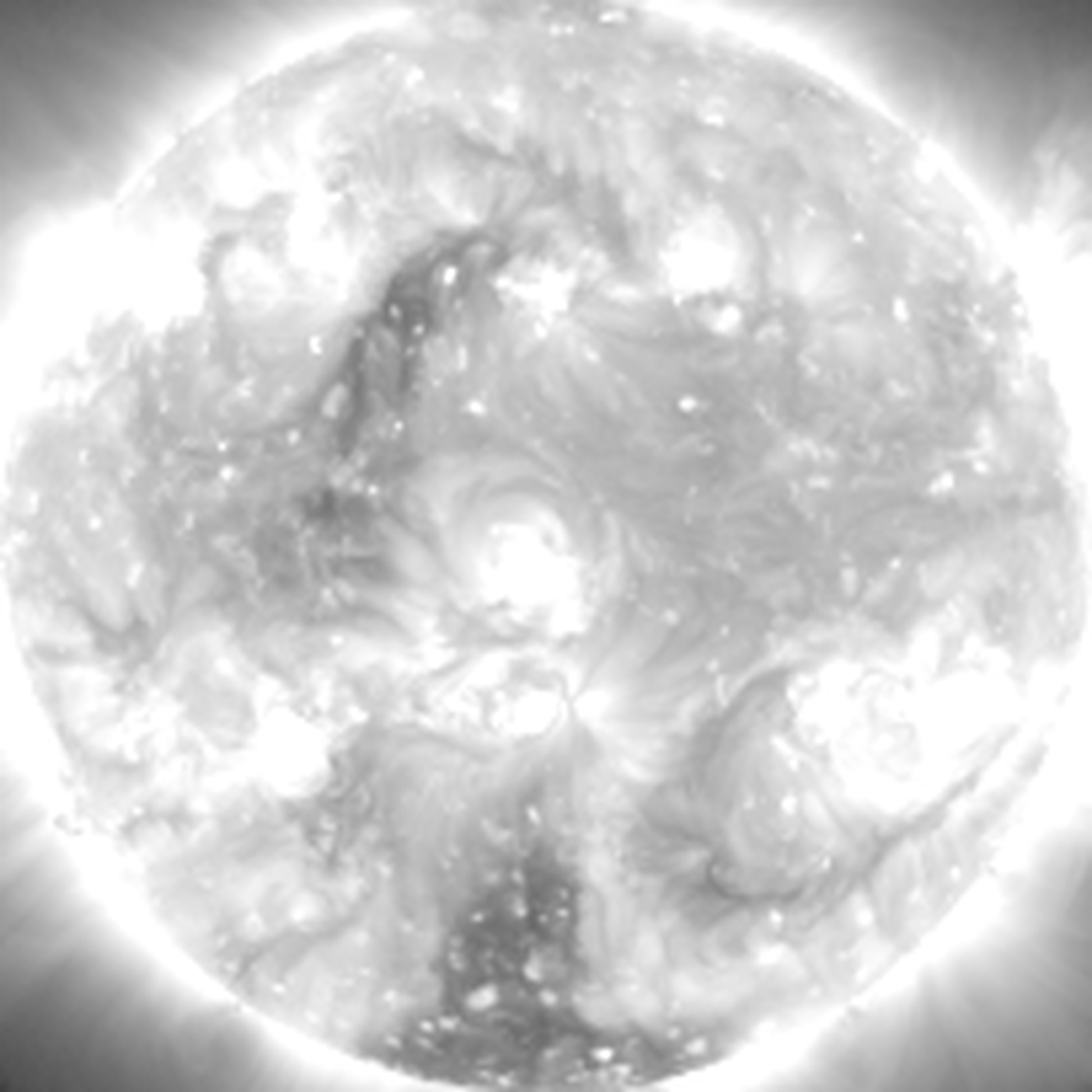}
\caption{Processed AIA.}
\end{subfigure}
\hfill
\begin{subfigure}{0.32\textwidth}
\centering
\includegraphics[width=\textwidth]{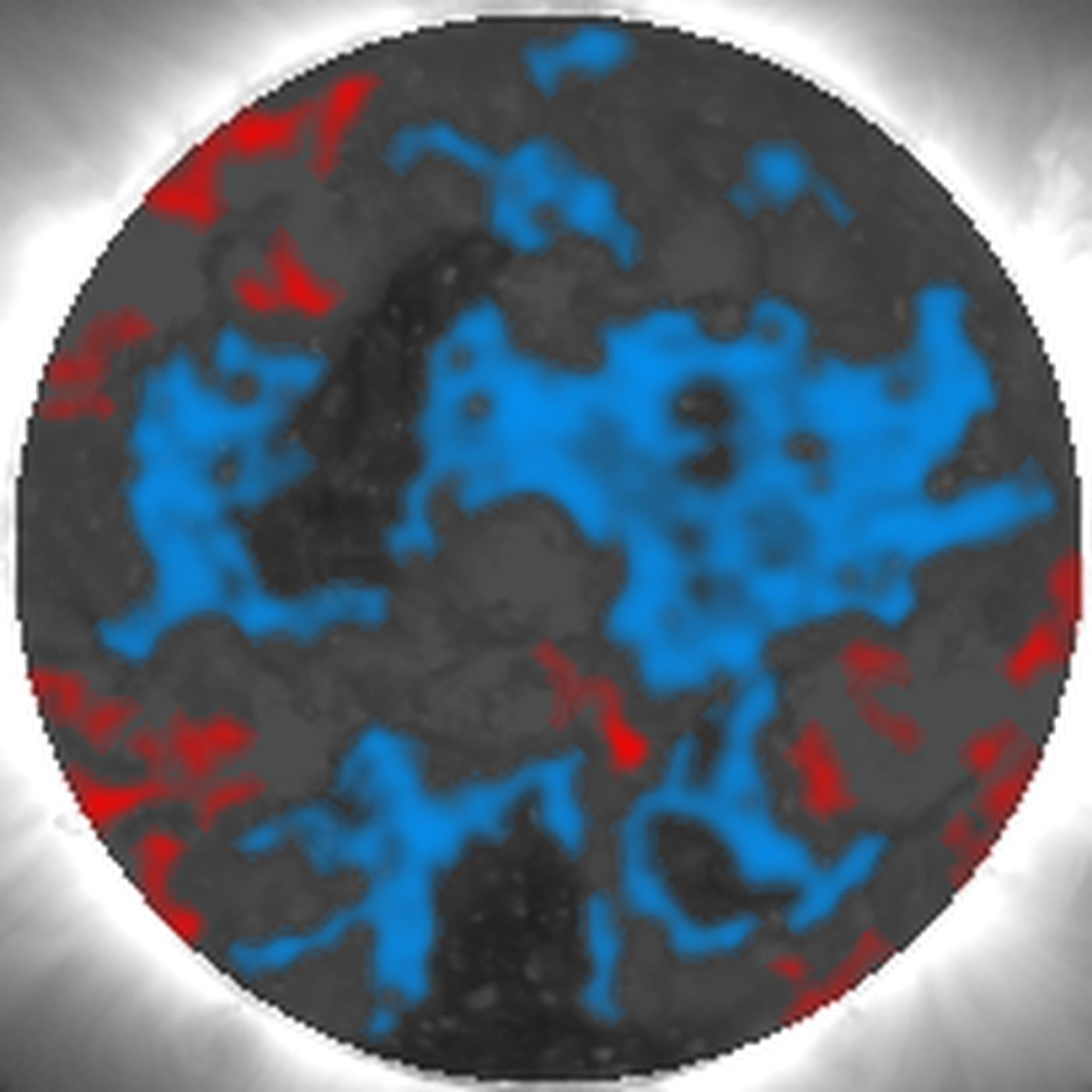}
\caption{$H_0$ evidence.}
\end{subfigure}
\hfill
\begin{subfigure}{0.32\textwidth}
\centering
\includegraphics[width=\textwidth]{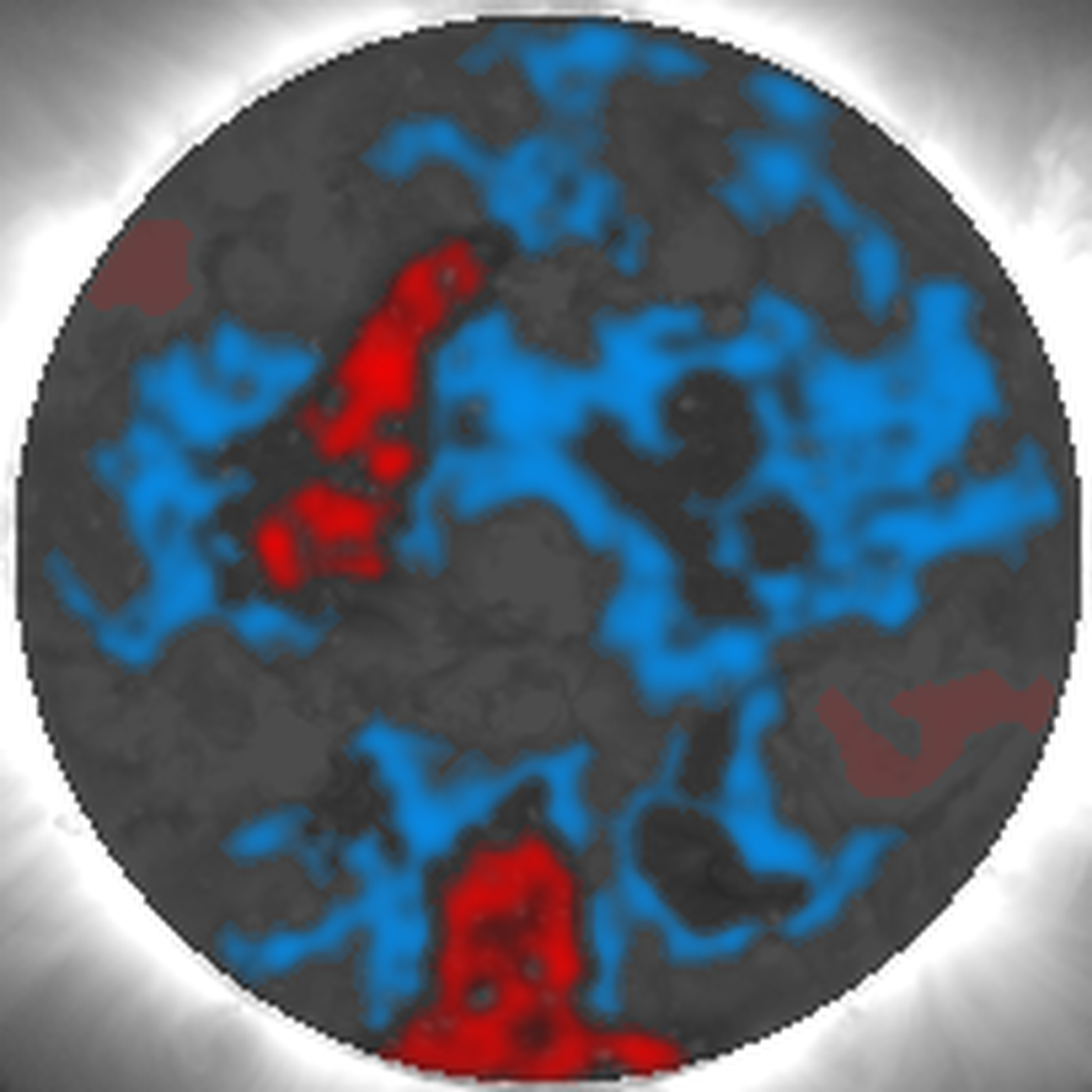}
\caption{$H_1$ evidence.}
\end{subfigure}
\caption{False negative example (2022-02-16 21:00): processed AIA view with corresponding $H_0$ and $H_1$ projected evidence maps. Red indicates HSS-supporting projected evidence, blue indicates non-HSS-supporting projected evidence.}
\label{fig:false_negative}
\end{figure}

\begin{figure}[h]
\centering
\begin{subfigure}{0.32\textwidth}
\centering
\includegraphics[width=\textwidth]{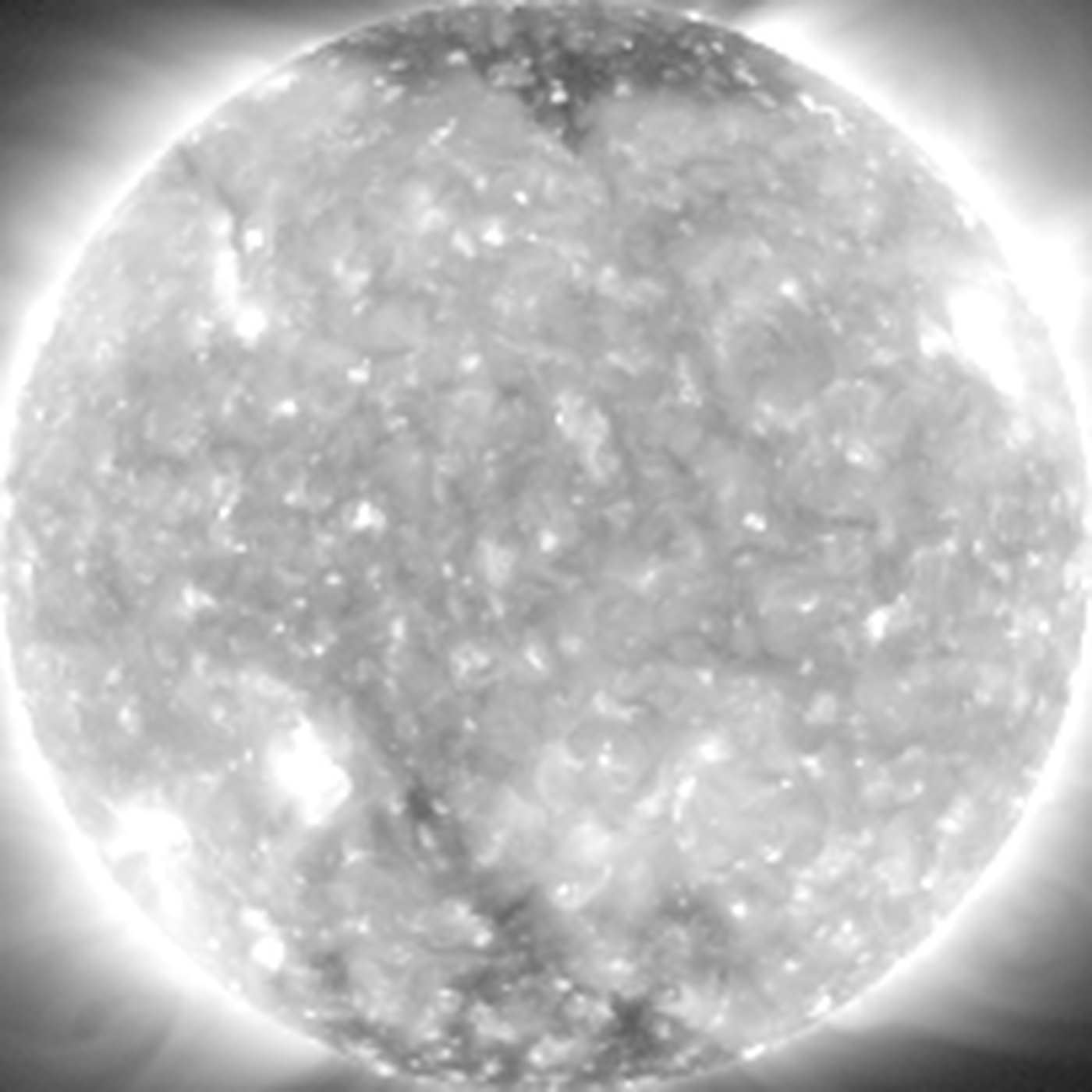}
\caption{Processed AIA.}
\end{subfigure}
\hfill
\begin{subfigure}{0.32\textwidth}
\centering
\includegraphics[width=\textwidth]{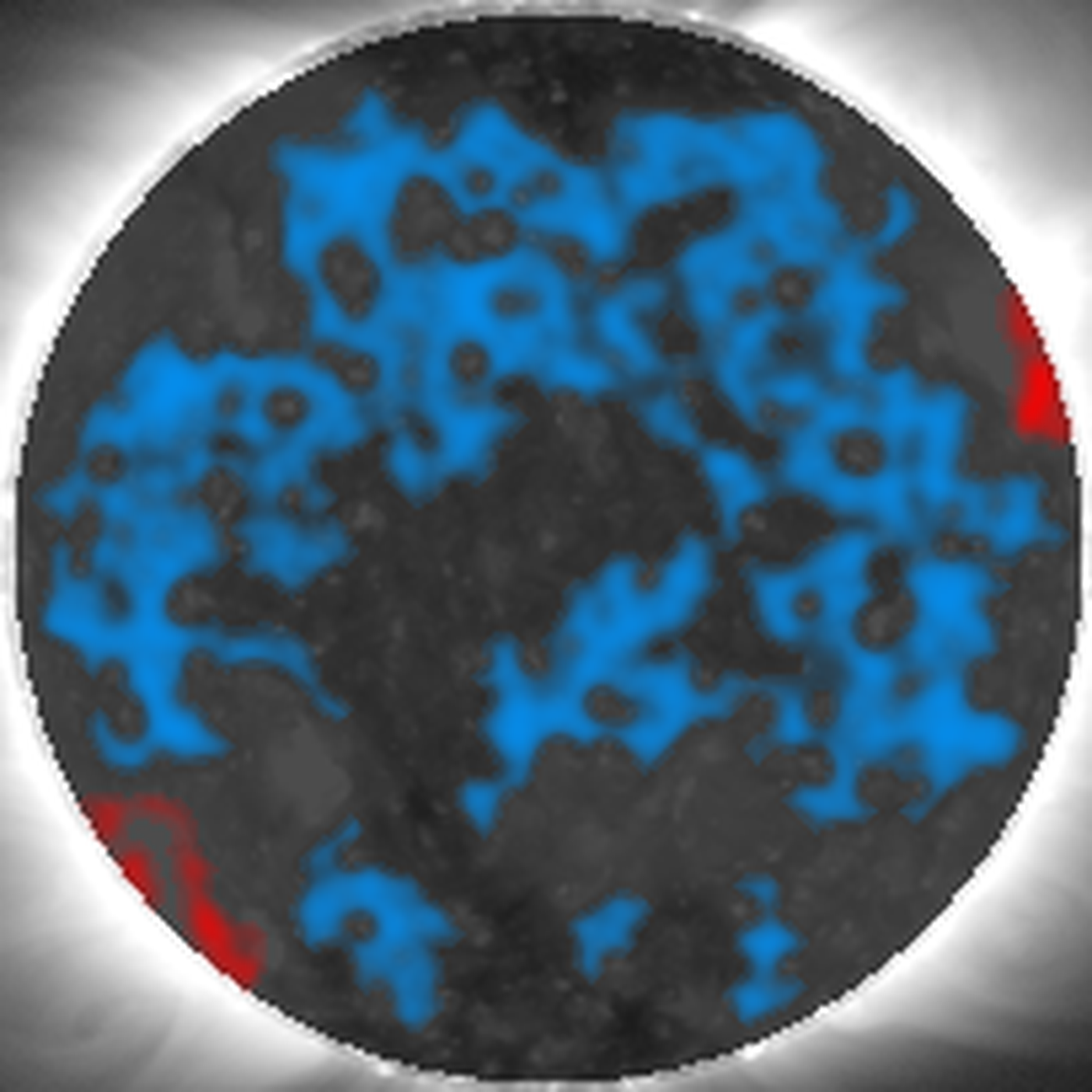}
\caption{$H_0$ evidence.}
\end{subfigure}
\hfill
\begin{subfigure}{0.32\textwidth}
\centering
\includegraphics[width=\textwidth]{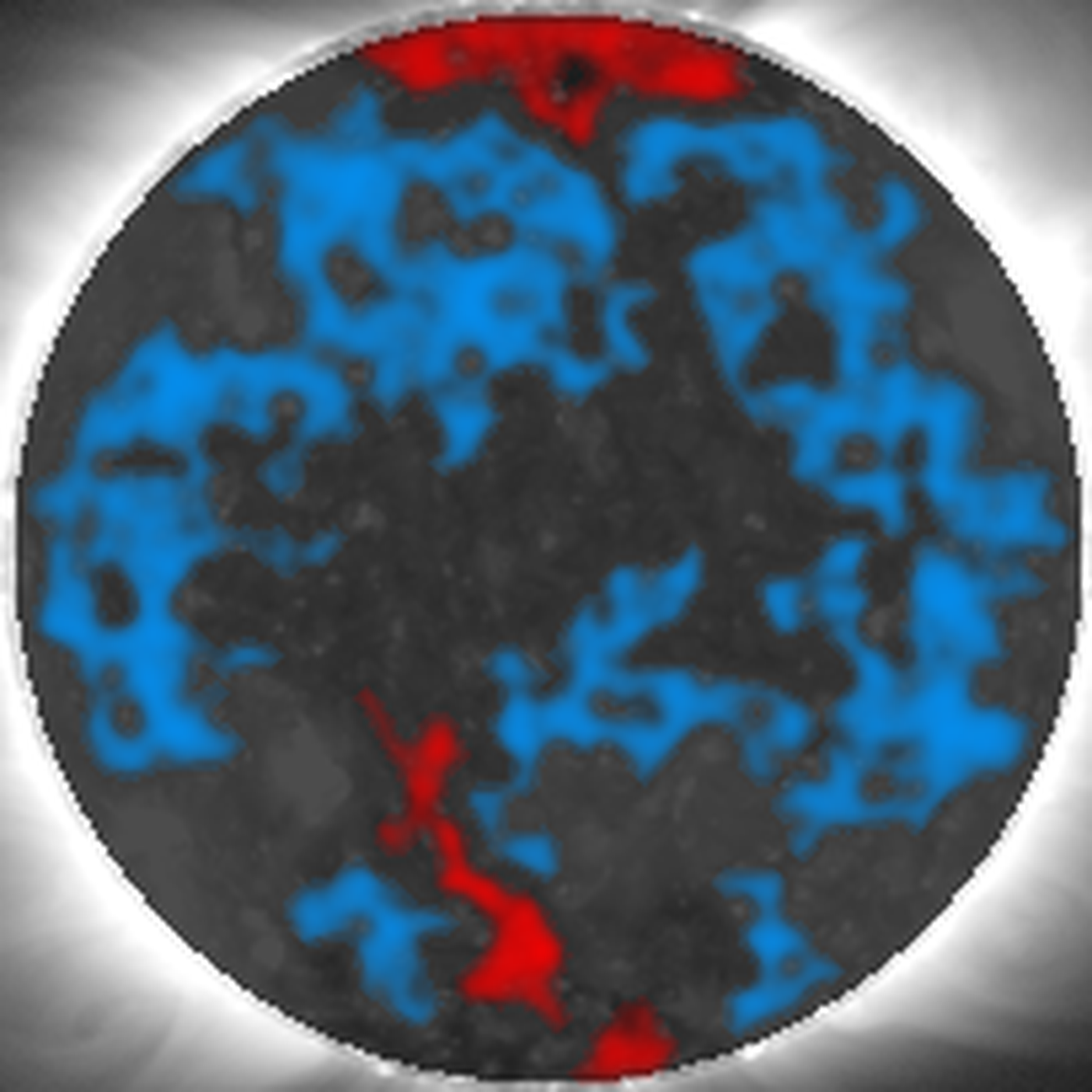}
\caption{$H_1$ evidence.}
\end{subfigure}
\caption{False positive example (2020-08-14 09:00): processed AIA view with corresponding $H_0$ and $H_1$ projected evidence maps. Red indicates HSS-supporting projected evidence, blue indicates non-HSS-supporting projected evidence.}
\label{fig:false_positive}
\end{figure}

In addition to the HSS example presented in the previous subsection, here we include in the discussion the topological annotation linked to a correctly classified non-HSS sample in Figure
\ref{fig:quiet}, a false negative sample in Figure \ref{fig:false_negative} and a false positive one in Figure \ref{fig:false_positive}.

Across the four examples, a first qualitative point is that HSS-supporting (red) evidence is present in all cases. This is expected, and it also explains why a purely visual ``red area vs blue area'' reading is not sufficient. The classifier decision is made in persistence-image feature space (see equation \eqref{eq:linear_model}), not in the displayed overlay space. Therefore, what changes the class is the sign of the signed weighted sum over all bins, including many weak contributions that are not visually salient in the final maps. The red/blue overlays are a post-hoc projection of those signed contributions (Section \ref{sec:xai}), further processed for readability by persistence-axis collapse, smoothing, quantile onset ($q=55\%$), and separate red/blue normalization. Consequently, they are faithful as qualitative localization maps, but not as a literal visual encoding of the exact decision process. For reference, the stored LogisticRegression+image probabilities for these examples are: $p\approx 1.0$ for the HSS true positive (Figure \ref{fig:storm}), $p=9.48\times10^{-10}$ for the non-HSS true negative (Figure \ref{fig:quiet}), $p=0.579$ for the selected false positive (Figure \ref{fig:false_positive}), and $p=0.459$ for the selected false negative (Figure \ref{fig:false_negative}).

We highlight that our model effectively associates HSS-supporting regions in $H_1$ to CHs, strongly indicating that the topological analysis was able to extract physically relevant information. Moreover, to test whether the slight HSS-supporting contribution near active-region boundaries is recurrent, we performed an exploratory post-hoc analysis of the 30 EUV images for which Logistic Regression models based on persistence images had been used to generate spatial evidence maps for annotation. These images represent a selected subset rather than the entire test set.
For each image $t$, we defined a positive-evidence mask $P_t$ as the union of red pixels in the $H_0$/$H_1$ evidence maps. 
We also defined an approximate AR-boundary mask $A_t$ from the corresponding normalized AIA image. Specifically, we identified the brightest regions within the solar disk and applied morphological operations to extract their boundaries. This mask therefore provides a brightness-based proxy for AR boundaries rather than an explicit segmentation of individual active regions.
To quantify the spatial association between positive evidence and these boundaries, we computed
\[
r_t = \frac{|P_t\cap A_t|}{|P_t|}, \qquad a_t = \frac{|A_t|}{|D_t|}, \qquad e_t = \frac{r_t}{a_t},
\]
where $D_t$ is the disk mask.
Here, $r_t$ is the fraction of positive-evidence pixels overlapping the approximate AR boundaries, while $a_t$ is the fraction of the solar disk occupied by those boundaries. Their ratio, $e_t$, measures the spatial enrichment of positive evidence near AR boundaries relative to a uniform distribution over the solar disk. Values $e_t>1$ indicate greater overlap than expected under this reference distribution.
We find $e_t>1$ in 24/30 targets (80\%), with median enrichment factor $1.46$ (interquartile range $1.05$--$1.81$). 
However, the spatial overlap was generally limited: the median fraction of positive-evidence pixels overlapping the AR-boundary proxy was $5.9\%$, exceeding $10\%$ in only $9$ of the $30$ images. These findings suggest a recurrent but relatively weak spatial association between positive topological evidence and bright-region boundaries in the selected images. 

This spatial association should be interpreted with caution, as it may reflect both properties of the underlying solar structures and characteristics of the explainability procedure. From a physical perspective, the complex intensity patterns surrounding active regions may contribute to the topological information extracted from EUV images. However, a methodological explanation is also possible. In constructing the spatial evidence maps, feature contributions in the persistence-image representation are aggregated along the persistence dimension to obtain a profile as a function of birth level. Consequently, multiple low-persistence topological features with similar birth values may collectively produce a noticeable contribution, even when their individual contributions are small. The positive evidence observed near bright-region boundaries may therefore partly reflect the accumulation of weak, short-lived topological features rather than the contribution of highly persistent structures. Overall, these findings suggest a recurrent but secondary spatial association, which should not be interpreted as evidence that active-region boundaries independently predict HSS occurrence.

\subsection{Discussion of limitations}
Our approach prioritizes interpretability and reproducibility, but several limitations should be kept explicit.

First, the explanation is \emph{global-model based}: projected maps are derived from fold-aggregated weights and sample features, not from a local surrogate fitted around a single image. This is useful for identifying recurrent discriminative patterns, but it does not by itself establish local causal mechanisms for one specific event.

Second, label construction is intentionally strict ($>500$ km/s for positive samples, i.e HSS samples, $<400$ km/s for negative samples, i.e. non-HSS samples). This improves class separability and interpretability of decision boundaries, but removes intermediate regimes that may still be physically meaningful. The reported metrics therefore describe performance under this strict operational definition, not under a fully continuous wind-speed target.

Third, although the temporal cross-validation strategy reduces the risk of information leakage associated with the recurrence of solar structures over successive rotations, model performance varies across the different temporal folds. Nevertheless, the results consistently indicate that linear classifiers combined with persistence images provide promising predictive performance. Small differences in performance between individual model configurations should therefore be interpreted with caution, particularly when they are comparable to the variability observed across folds.

Fourth, the image projection is not unique. We adopt birth-axis aggregation because it provides a deterministic and interpretable mapping from feature space to image space: each pixel has a direct local scalar value, hence a natural birth-bin assignment, while death/persistence coordinates generally do not admit a unique local pixel assignment without additional cycle-tracking assumptions. This collapse therefore improves interpretability and reproducibility of the projection operator. By contrast, avoiding the aggregation and projecting full birth--persistence contributions would require extra modeling assumptions (e.g., local allocation rules for persistence/death, generator tracking, or kernel redistribution). Those alternatives can preserve more 2D persistence information, but they are less direct, more sensitive to implementation choices, and harder to compare across samples.

Finally, the pipeline remains topology-only by design. This is a strength for transparency, but it may miss complementary geometric descriptors (e.g., explicit CH area/elongation) that could improve calibration in edge cases. Particularly, previous studies (see, e.g. \citet{rotter2012,reiss2016,collin2025}) have found that the main CH properties that contribute to the solar wind speed are the CH area and the location of the CH. These properties are not encoded by the topological features. A controlled hybrid benchmark is thus a natural next step rather than a contradiction of the present framework.

\section{Conclusions}\label{sec:conclusions}
In this work, we investigated whether topological information extracted from full-disk EUV solar images contains predictive information on the occurrence of coronal-hole-driven high-speed solar wind streams at Earth four days later. By combining persistent homology with different vectorization techniques and machine-learning classifiers, we showed that the topology of EUV intensity patterns provides measurable predictive information, even without explicitly identifying or segmenting individual coronal holes or introducing predefined geometrical descriptors.

Among the investigated approaches, persistence images combined with linear classifiers provided particularly promising results, achieving an AUC of approximately 0.82 and a TSS of approximately 0.42--0.43. These findings suggest that the connectivity and hole structure of EUV intensity patterns, characterized across multiple filtration levels, contain information relevant to the subsequent occurrence of HSSs at Earth.

Beyond classification performance, we developed an explainability framework that connects the topological representations to the classifier decisions and their spatial interpretation in the original EUV observations. By examining both global feature relevance and image-specific contributions, the proposed approach allows us to identify which regions of the topological representation influence the classification and to explore how the corresponding evidence is distributed across the solar disk. The worked example illustrates the complete connection between persistence diagrams, vectorized features, model coefficients, and spatial evidence maps, providing a means to investigate the relationship between topological information and observed coronal structures.

The present study is intended as an assessment of the predictive information contained in EUV image topology rather than as an operational HSS forecasting system. Its conclusions are limited by the adopted binary event definition, the fixed four-day prediction horizon, and the dependence of the spatial evidence maps on the projection procedure. Future work will investigate the sensitivity of the results to different prediction horizons and HSS definitions, assess the robustness of the spatial attribution maps, and compare topology-based representations with approaches combining topological and geometrical information under the same temporal validation framework.

\begin{authorcontribution}
F.M. developed and implemented the methodology and performed the analysis. D.C. contributed to dataset preparation. S.G. and D.C. contributed to the experimental design and contributed to the interpretation of the results. F.M. wrote the main manuscript text. All authors reviewed and approved the manuscript.
\end{authorcontribution}

\begin{fundinginformation}
AIA data used here are available by courtesy of the AIA and HMI consortia.
We acknowledge use of NASA/GSFC's Space Physics Data Facility's OMNIWeb service.
F.M. and S.G. acknowledge financial support from the Istituto Nazionale di Astrofisica (INAF) through the grant “Investigating coronal holes' shapes and their relationship with high-speed solar wind streams via topological data analysis”. F.M. and S.G. are also grateful to the Gruppo Nazionale per il Calcolo Scientifico - Istituto Nazionale di Alta Matematica (GNCS - INdAM).
D.C. acknowledges the support of the Helmholtz International Berlin Research School in Data Science (HEIBRiDS).
\end{fundinginformation}

\begin{dataavailability}
The observational data and HSS list used in this study are the same as those described in \cite{collin2025} and are publicly available through \url{https://doi.org/10.5880/gfz.2.7.2024.001} \cite{Collin2024data}. Software for downloading and preprocessing of solar images was developed in \cite{collin2026} and is available at \url{https://github.com/DanielCollin96/solar_image_processing} \cite{Collin2026solarimagecode}.
\end{dataavailability}

\bibliographystyle{spr-mp-sola}
\bibliography{references}

\end{document}